\documentclass[lettersize,journal]{IEEEtran}
 
\usepackage{mathtools}
\usepackage{balance}
\usepackage{ amssymb }
\usepackage{listings}
\usepackage{graphicx}
\usepackage{enumitem}
\usepackage{caption}
\usepackage{subcaption}
\usepackage{multirow}
\usepackage[hidelinks = true]{hyperref}
\usepackage[linesnumbered,ruled,vlined,algo2e]{algorithm2e}
 
 \SetCommentSty{mycommfont} 
\usepackage{tikzscale}
\usepackage{tikz}
\usetikzlibrary{calc}
\usepackage{array}
\usetikzlibrary{shapes}
\usetikzlibrary{positioning}
\usetikzlibrary{arrows.meta}
\usepackage{circledsteps}
\usepackage{graphicx}
\usepackage{mathtools}
\usepackage{svg}
\definecolor{siamgreen}{HTML}{159a88}
\tikzset{
blueNode/.style={circle,draw=black!50,fill=blue!20,thick,
inner sep=0pt,minimum size=3mm},
orangeNode/.style={circle,draw=black!50,fill=orange!80,thick,
inner sep=0pt,minimum size=3mm},
redNode/.style={circle,draw=black!50,fill=red!80,thick,
inner sep=0pt,minimum size=3mm},
}

\newcommand\circled[1]{\tikz[baseline=(char.base)]{
            \node[shape=circle,fill=black,inner sep=0.25pt] (char) {\textcolor{white}{\sffamily\small #1}};}}

\definecolor{deepgreen}{rgb}{0,0.7,0}

\definecolor{ssgreen}{HTML}{138808}

\newcommand{\ours}{{\texttt{BANG}}\xspace} 

\newcommand{\banginmemory}{\hbox{\texttt{BANG In-memory}}\xspace}
\newcommand{\bangexactdist}{\hbox{\texttt{BANG Exact-distance}}\xspace}
\newcommand{\diskann}{{\texttt{DiskANN}}\xspace}
\newcommand{\vamana}{{\texttt{Vamana}}\xspace}
\newcommand{\mypara}[1]{\noindent\textbf{#1.}}

\newcommand{\revchanges}[1]{{\color{black}{#1}}} 
\newcommand{\revonechanges}[1]{{\color{black}{#1}}} 
\SetAlgoNlRelativeSize{0}
\SetAlgoNlRelativeSize{-1}
\SetAlgoNlRelativeSize{-2}
\newcommand{\coloredtcp}[1]{\textcolor{gray}{\textit{\footnotesize// {#1}}}} 
\newcommand{\coloredtcc}[1]{\textcolor{gray}{\textit{\footnotesize/* {#1} */}}} 
\newcommand{\speedupatpointnine}{{50$\times$-400$\times$}\xspace}

\begin{document}
    \title{{\normalfont \ours}: Billion-Scale Approximate Nearest Neighbour Search using a Single GPU} 


\author{Karthik V., Saim  Khan, Somesh Singh, Harsha Vardhan Simhadri and Jyothi  Vedurada
 \thanks{$\bullet$ Karthik V., Saim  Khan and Jyothi Vedurada are with IIT Hyderabad, India.
 \textit{E-mail: cs21resch14001@iith.ac.in, jyothiv@cse.iith.ac.in.}
 }
 \thanks{$\bullet$ Harsha Vardhan Simhadri is with Microsoft, USA.} 
 \thanks{$\bullet$ Somesh Singh is with LabEx MILYON and LIP (UMR5668), France.}
}

\maketitle


	\begin{abstract}

 Approximate Nearest Neighbour Search (ANNS) is a subroutine in algorithms routinely employed in information retrieval, pattern recognition, data mining, image processing, and beyond. 
Recent works have 
established that
graph-based ANNS algorithms 
are
practically more efficient 
than the other methods proposed in the literature.
The growing volume and dimensionality of data
necessitates designing \textit{scalable} techniques for ANNS. 
To this end, the prior art has explored parallelising graph-based ANNS on GPU, leveraging its 
massive parallelism.
The current state-of-the-art GPU-based ANNS algorithms either (i) require both the dataset and the generated graph index 
to reside entirely in the GPU memory, 
or (ii) they partition the dataset into small independent \textit{shards}, each of which can fit in GPU memory, and 
perform the search on these shards on the GPU. 
While the first approach  
fails to handle
large datasets  
due to the limited memory available on the GPU, the latter
delivers
poor performance on large datasets due to high data traffic over the low-bandwidth PCIe interconnect. 



We introduce \ours, a first-of-its-kind 
technique for
graph-based ANNS on GPU
for billion-scale datasets, 
that cannot entirely fit in the GPU memory.
\ours stands out by harnessing a compressed form of the dataset on a single GPU to perform distance computations while efficiently accessing the graph index kept on the host memory, enabling efficient ANNS on large graphs within the limited GPU memory.
\ours incorporates highly optimised GPU kernels and 
proceeds in
phases that 
run
concurrently
on the GPU and CPU, taking advantage of their architectural specificities. 
Furthermore, 
it enables overlapping communication with computation that results in efficient data transfer between the CPU and GPU.
\revonechanges{
Using a single NVIDIA Ampere A100 GPU, \ours achieves throughputs \speedupatpointnine  higher than competing methods 
for a recall of 0.9 on three popular billion-scale datasets.
}
%
	\end{abstract}

\section{Introduction}
\label{s:intro}
The $k$-Nearest-Neighbour-Search ($k$NN) problem is to find the $k$ nearest data points 
to a given query point in 
a multidimensional dataset.
As the dimensionality increases, 
\textit{exact}
search methods 
become increasingly inefficient.
In order to evaluate the exact $k$-nearest-neighbours of a query point in a $d$-dimensional dataset having $n$ points, as a consequence of the \textit{curse of dimensionality}, all $n$ points must be examined; this takes $O(nd)$ time~\cite{indyk}.
Therefore, it has become commonplace to use Approximate Nearest Neighbour Search (ANNS) for finding the nearest neighbours of a query, to mitigate the curse of dimensionality by sacrificing a small accuracy for speed~\cite{GTS}.
\revchanges{ANNS accuracy is typically measured as \textit{recall}, which quantifies the overlap between the retrieved nearest neighbours and the ground truth for a query.
}
ANNS is widely applied in information retrieval~\cite{drugdiscovery, genomesequencing, deepchemistry}, recommendation systems~\cite{GenAIRecommender,alibaba}, search engines~\cite{bingsearch}, and computer vision~\cite{SegmentGraph} to search through large datasets of words, documents, images, and multimedia. 
Advances in deep learning~\cite{deeplearningAI} have made ANNS essential for similarity search~\cite{GTS}.
ANNS is a key subcomponent of vector databases~\cite{vectorydatabasesurvey} and is transforming the way vector embeddings are efficiently indexed and queried~\cite{Milvus_blog}. 
Search queries on such massive multidimensional datasets are often processed in batches to meet 
the high throughput demands, as evidenced by recent efforts~\cite{ faiss, ggnn, GANNS, song}  leveraging the massive parallelism of GPUs for ANNS.
One of the earliest use cases of GPUs in this context was accelerating $k$NN queries on road networks, such as locating the $k$ nearest cars for ride-sharing~\cite{gpuroadnetworks}. 
\revonechanges{
Nowadays, GPUs are increasingly powering vector databases for Retrieval-Augmented Generation (RAG)~\cite{nvidia_rag}, boosting semantic accuracy and freshness of results in static Large Language Models (LLMs) and supporting high-throughput use cases, such as cloud-based chatbots~\cite{vectordatabse_chatbot} and unstructured data applications like e-commerce~\cite{vectordatabse_ecommerce}.
}



Billion-scale ANNS is computationally intensive, with pairwise distance calculations well-suited for GPU architectures.
Zhang et al.~\cite{rummy} compare CPU- and GPU-based ANNS techniques, and conclude that GPUs offer higher performance and cost-effectiveness for accelerating vector query processing.
The Billion-Scale Approximate Nearest Neighbor Search Challenge~\cite{neuripscompetition_report} evaluates ANNS on various hardware platforms (e.g. CPU, GPU, custom hardware) and establishes that GPU-based implementations significantly outperform. 
Billion-scale datasets exceed single GPU memory, and hence ANNS on these datasets requires multiple GPUs~\cite{ggnn}; this introduces two challenges:
(a) a significant escalation in implementation cost (as seen in the Track3 Cost/Power Leaderboards~\cite{neuripscompetition_report}),
(b) a mismatch between GPU compute performance and memory resources.

Graph-based ANNS algorithms~\cite{hnsw, nsg, diskann} have been shown to be generally more efficient, in practice, at handling large datasets. However, GPU implementations of these algorithms~\cite{cagra,song,ggnn}
require storing graph data structures in GPU memory, which limits their ability to handle large datasets.
Even 
recent GPUs like the NVIDIA Ampere A100, having 80GB device memory, cannot accommodate the entire input data (graph index and data points).
Prior solutions, such as sharding, effectively implemented by GGNN~\cite{ggnn}, incur high memory transfer cost. 
For example, with the PCIe 4.0 interconnect operating at its peak theoretical transfer rate of 32 GB/s, transferring the DEEP1B dataset (having a data size of 384GB and a graph index size of 260GB) from CPU to GPU would take 20 seconds. This would result in a low throughput of 
less than
500 QPS for 10,000 concurrent queries on the GPU.
Alternatively, hashing and compression techniques, as showcased by SONG~\cite{song} and FAISS~\cite{faiss}, can handle 
large volumes of data using a single GPU by reducing the data dimensionality or by compressing the vectors. 
However, these approaches 
face limitations in achieving high recall on massive datasets.
It has been shown that it is feasible to attain high throughputs at high recalls on large datasets using multiple GPUs~\cite{ggnn}. 
However, such approaches have a very high hardware cost which makes them undesirable.
Thus, this paper explores an important question: 
\emph{Can we increase the throughput of ANNS queries without compromising their recall by using a single GPU?} 

In this paper, we introduce \ours{}, a novel GPU-based ANNS method, which is 
designed to efficiently handle large datasets, with billions of points, that cannot entirely fit in GPU memory. 
\ours{} stands out by employing compressed data for accelerated distance computations on the GPU while keeping the graph index and the dataset on the CPU, 
thus enabling efficient ANNS on large datasets.
CPU transfers neighbour information to the GPU on demand in every iteration of the search.
Furthermore, in contrast to the previous approaches that treat search activity as one monolithic block, \ours divides ANNS activity into distinct phases, 
in order to maximise
resource utilisation and 
mitigate
CPU-GPU data transfer bottlenecks. 
Specifically, the division into phases leads to three advantages:
(1)~executing different phases on CPU and GPU in a pipeline
(2)~optimising each phase's (GPU kernel's) span characteristics separately to maximise parallelism on GPU
(3)~asynchronously prefetching required data from CPU to GPU for 
specific
phases.
Thus, the overall performance of \ours stems from several factors, such as CPU-GPU load balancing, highly optimised GPU kernels (for tasks such as distance calculations, sorting, and updating worklists), prefetching and pipelining.

ANNS methods typically follow the \textit{index} and \textit{search} paradigm whereby data points are first processed to construct an efficient index, and search queries are then processed using this index.
In this work, our focus is efficient billion-scale ANNS on GPU utilising an underlying graph index, and therefore we do not construct a graph index but instead utilise an existing one from prior work~\cite{diskann}.

\revchanges {
Our evaluations show that 
\ours
significantly outperforms the state-of-the-art on four popular real-world billion-size ANNS  datasets with varying characteristics across all recall values on a single NVIDIA A100 GPU. 
\revonechanges{
Notably, on the SIFT-1B and Deep-1B datasets, we achieve $50.5 \times$ higher average throughput over competing methods at a high recall of 0.95.}
}




 This paper makes the following main contributions:
\begin{itemize}
\item We introduce \ours, a novel GPU-based ANNS method for efficiently searching billion-scale datasets using a single GPU.
\item We propose a \textit{phased-execution} strategy that enables pipelined execution of ANNS steps on both CPU and GPU, optimising CPU-GPU load balancing, asynchronous data transfer, and GPU kernel performance for maximising parallelism. 
\item We present 
efficient
GPU kernels for distance calculations and worklist updates in ANNS and implement prefetching and pipelining to effectively utilise CPU and GPU, reducing traffic over PCIe interconnect and enhancing throughput while retaining 
recall.
\item We conduct an extensive evaluation demonstrating that \ours significantly outperforms state-of-the-art GPU-based ANNS methods in terms of throughput and cost.
\end{itemize}


\section{Background}
\label{s:back}


\subsection{GPU Architecture and Programming Model} 
\label{ss:gpu_arch}
Graphic Processor Units (GPUs) are accelerators that offer massive multithreading and high memory bandwidth~\cite{nvidiacuda}. 
CPU (\textit{host}) and GPU (\textit{device}) are connected via an interconnect such as PCI Express that supports CPU$\rightleftarrows$GPU communications.
For our implementation, we use NVIDIA GPU with the CUDA~\cite{cuda} programming model.
The main memory space of GPU is referred to as \textit{global memory}.
A GPU comprises several CUDA cores, which are organised into \textit{streaming multiprocessors} (SMs). 
All cores of a SM share on-chip \textit{shared memory}, which is a software-managed L1 data cache. 
GPU procedures, referred to as \textit{kernels}, run on the SMs in parallel by launching a \textit{grid} of \textit{threads}.
Threads are grouped into \textit{thread-blocks}, with each block assigned to a SM. 
Threads within a thread-block communicate and synchronise via shared memory. 
A thread-block comprises \textit{warps}, each containing 32 threads, executing in a single-instruction-multiple-data (SIMD) fashion.
Kernel invocation and memory transfers are performed using task queues called \textit{streams}, which can be used for task parallelism. 
\subsection{Graph Index }\label{s:backdiskann}
Several of the best-performing  
ANNS algorithms are graph-based, achieving both high recall
and high throughput, measured as queries per second (QPS).
A graph-based ANNS algorithm runs on a \textit{proximity graph}, which is pre-constructed over the dataset points by connecting each point to its nearby points. 
Our \ours{} algorithm implements 
an efficient ANNS method that can run on various proximity graphs, such as kNN~\cite{ggnn}, HNSW~\cite{hnsw}, Vamana~\cite{diskann}, and others. 
Manohar et al.~\cite{ParlayANN} conduct a study on various CPU-based ANNS algorithms and show that  the \textit{DiskANN}~\cite{diskann} approach, which runs on the Vamana graph index, achieves superior throughput and recall on large datasets.
\revchanges{Furthermore, the recent work by Wang et al.~\cite{starling}, which proposes enhancements to the Vamana graph index for disk-resident indexes, also highlights the significance of the Vamana graph.}
Hence, we employ the Vamana graph as the underlying graph index 
in the implementation of \ours{}.
Vamana graph index is constructed by sharding the large dataset into smaller overlapping clusters.
The technique iteratively processes individual clusters and uses \textit{RobustPrune} and \textit{GreedySearch} routines to construct directed subgraphs on the host RAM, which are finally written to an SSD.
The final graph is formed by merging the edges in the subgraphs. 
While prior techniques that construct the graph index using the SNG property~\cite{SNG-10.5555/313559.313768} aim to reduce distances to the query point along the search path, Vamana further reduces the number of hops to the query point with the RobustPrune property, resulting in a faster search.
\section{ANN Search on GPU: Challenges in Handling Billion-scale Data}
\label{s:overview}


We propose \ours{}, a method for parallel  
ANNS on a single GPU.
There are two main challenges in parallelising ANNS on a single GPU:
(1) managing large graphs within the constraints of limited GPU memory, (2) extracting sufficient parallelism for ANNS to fully utilise the hardware resources.
In the following subsections, we discuss these challenges in detail. 

\subsection{Limited GPU Memory}
\label{subsec:limited_memory}
 A primary challenge when dealing with large datasets on GPUs is their massive memory footprint.
Table~\ref{tbl:data} shows the sizes of various billion-scale ANNS datasets and their respective graph sizes.
Notably, even with the recent 
A100 GPU's maximum global memory capacity of 80GB, it is evident that the entire input data (base, graph, and query) cannot be accommodated within the GPU memory.
To address this challenge, 
a recent work GGNN~\cite{ggnn} effectively implemented sharding as a solution. 
However, this approach requires frequent swaps of both processed and unprocessed shards between GPU and host RAM, resulting in high memory transfer costs over the PCI interconnect. 
Further, these swaps may result in an idle time of CPU/GPU, and optimal GPU-CPU coordination is essential during iterative graph traversals of ANNS.
Furthermore, sharding the entire data is not viable because 
even with the PCIe 4.0 interconnect operating at its peak theoretical transfer bandwidth of 32 GB/s, 
it will take an estimated 20 seconds to transfer the largest dataset from CPU to GPU and will result in a markedly low throughput of less than 500 QPS as explained in Section~\ref{s:intro}.

Although employing a multi-GPU setup can facilitate the effective distribution of shards across all GPUs to store the entire input data (as demonstrated with eight GPUs in GGNN~\cite{ggnn}), this approach incurs significant hardware cost. 
As a result, our focus is on developing an efficient and cost-effective parallel implementation of ANNS using a single GPU.

\begin{table}[t]
\centering
\caption{Sizes of Data and Graphs for Various Datasets.}
\scalebox{0.9}{
\begin{tabular}{@{~~}l|r|r|r}
 \multicolumn{1}{c|}{\textbf{Dataset}} & \textbf{Data Size} & \textbf{Vamana Graph}~\cite{diskann} & \textbf{kNN Graph}~\cite{ggnn}  \\\hline
SIFT-1B & 128 GB & 260 GB & 80 GB\\\hline
Deep-1B & 384 GB &  260 GB & 96 GB \\\hline
MSSPACEV-1B & 100 GB  & 260 GB & 80 GB \\\hline

\revchanges{Text-to-Image-1B} & \revchanges{800 GB}  & \revchanges{260 GB} & \revchanges{80 GB} \\
\end{tabular}
}
\label{tbl:data}
\end{table}
Alternatively, hashing and quantization techniques can handle large data on a single GPU.
Hashing effectively reduces data dimensionality to make it fit within GPU memory, as demonstrated by SONG~\cite{song}, which compresses a 784-byte vector to 64 bytes in the MNIST8M dataset. 
However, hashing is better suited for smaller datasets and may not achieve high recall on datasets with billions of points.
Data compression, as demonstrated by FAISS~\cite{faiss}, can accelerate query processing in GPU-based ANNS, but it cannot provide high recall.


To mitigate these limitations, we present a novel GPU-based solution wherein we conduct ANNS that does not require the large graph to be present in GPU memory. Instead, the CPU retrieves the required neighbourhood information from the graph index kept on the host RAM and transfers it to the GPU efficiently.
Likewise, since the GPU cannot accommodate the entire base dataset, the search process on the GPU employs compressed vectors (instead of the base dataset vectors) for distance calculations.
The compressed vectors are generated using the widely utilised Product Quantization~\cite{PQ} technique, employed in several previous works, including FAISS~\cite{faiss} and DiskANN~\cite{diskann}.
Thus, \ours{} alleviates the CPU-GPU data transfer bottleneck by transferring only compressed data to the GPU and eliminating the need to transfer the entire graph. Instead, it fetches neighbour information for a node from the CPU to the GPU on demand.


\subsection{Optimal Hardware Usage}
\ours{}'s approach of fetching neighbours from the CPU and calculating distances from them to the query on the GPU using compressed vectors presents two challenges.

The primary challenge is to prevent CPU and GPU idleness by ensuring their concurrent and continuous utilisation despite the dependence on neighbour information between these devices.
Furthermore,
it is important to consider the volume of data that needs to be transferred between the CPU and the GPU when balancing the work distribution between them to ensure that performance is not constrained by bandwidth limitations caused by prolonged data transfers, as in the GGNN~\cite{ggnn} framework.
To address these issues, we propose a parallel ANNS implementation \ours that maximises parallelism by efficiently utilising the hardware (CPU, GPU and PCIe bus) with a phased-execution approach to efficiently load balance the ANNS work across CPU and GPU, and to minimise CPU-GPU data transfer bottlenecks. 
The phased-execution strategy facilitates the execution of different steps of ANNS in a pipeline on both CPU and GPU, which enables: (1) CPU-GPU load balancing to minimise CPU and GPU idle time, (2) using prefetching techniques to transfer data from CPU to GPU asynchronously for certain phases, and 
(3) optimising the span characteristics of each GPU kernel (corresponding to operations such as distance computation, sorting and updating worklist) separately to maximise parallelism.

The second challenge is to optimise the placement of the data structure according to the computational strengths and memory access capabilities of the CPU and GPU, due to the higher memory footprint caused by thousands of queries running concurrently on the data in the GPU.
For example, when handling parallel queries (Q) on the GPU and managing data structures for tracking visited nodes (V) and their processing, in order to determine whether structures should be placed on the CPU or GPU, several key factors need to be considered, such as the computational complexity associated with the memory size of $Q \times V$, the patterns of memory accesses within V (access times vary with CPU and GPU memory bandwidths), and the time it takes to transfer data $Q \times V$ given the limited bandwidth of the PCIe interconnect between the CPU and GPU.
\ours{} handles this by using an efficient implementation of bloom filter~\cite{bloom70} on the GPU. 
We discuss this in detail in Section~\ref{ss:bloomfilter}.

\section{{\normalfont \ours}: Billion-scale ANNS on a Single GPU}
\label{s:kernels}

Given a set of available queries, $\mathcal{Q}$, \ours processes the queries in parallel by effectively using CPU (host), GPU (device) and PCIe (data transfer link between CPU and GPU), achieving high throughput on billion-scale data using a single GPU.
\ours{} is a graph-based ANNS technique.

\begin{algorithm2e}[t]
\DontPrintSemicolon 
\KwIn{A graph index $G$, a query $q$, required no. of neighbours $k$, medoid $s$ and a parameter $t$ ($\geq k$)}
\KwOut{A set $\{ c_1, c_2, \ldots, c_k \} $ of $k$ approx. nearest neighbours}
Initialise visited set $\mathcal{V}$ $\gets$ $\emptyset$ and worklist $\mathcal{L} \gets \{s\}$\;
\While{$\mathcal{L} \setminus \mathcal{V} \neq \emptyset$ 
}
{
  $u^*$ $\gets$ arg min$_{u\in \mathcal{L} \setminus \mathcal{V}}$ $\parallel$ x$_u$-x${_q}$ $\parallel$ \coloredtcp{get nearest candidate} \; 
  $\mathcal{V}$ $\gets$ $\mathcal{V}$ $\cup$ $\{u^*\}$\;
   $\mathcal{L}$ $\gets$ $\mathcal{L}$ $\cup$ $G.adj(u^*)$\;
  Update $\mathcal{L}$ to keep up to $t$ nearest candidates\;  
}
  $ \mathcal{K} \gets k$ candidates nearest to $q$ in $\mathcal{L}$\;
\Return{$\mathcal{K}$}\;

\caption{{\sc Greedy Search} }
\label{algo:BFS}
\end{algorithm2e}

In the case of graph-based ANNS algorithms~\cite{hnsw,song, diskann,ggnn}, it is a fundamental approach to use greedy search, depicted in Algorithm~\ref{algo:BFS}. 
A worklist $\mathcal{L}$ is used to hold at most $t$ entries (i.e. nodes in the graph) in the increasing order of the distances of the points from the query point.
The algorithm begins at a fixed point $s$ and explores the graph $G$ in a best-first order by evaluating the distance between each point in the worklist $\mathcal{L}$ and the query point $q$, advancing towards $q$ at each iteration and finally reporting $k$ nearest neighbours.

 \begin{algorithm2e}[t]

  \small
  \DontPrintSemicolon
  \caption{{\large \ours}: ANNS using a Single GPU. 
\label{alg:querylookup}}
      \KwIn{$G$, a graph index; $k$, required no. of nbrs; $s$, medoid  \newline   
      \textit{PQDistTable}, dist. b/w compressed vectors \& queries 
      \newline 
      $\mathcal{Q}_{\rho}$, a batch of $\rho$ queries, and a parameter $t$ ($\geq k$), size of the worklist that controls recall of the search
       }
      \KwOut{$\mathcal{K_\rho}\coloneqq\bigcup_{i=1}^{\rho} \{ \mathcal{K}_{i}$\}, where $\mathcal{K}_{i}$ is the  set of $k$-nearest neighbours for $q_i \in \mathcal{Q}_{\rho}$}

    \SetKwFunction{FetchNeighbors}{FetchNeighbors}
    \SetKwFunction{FilterNeighbors}{FilterNeighbor}
    \SetKwFunction{CalculateDistanceToQuery}{ParallelComputeDist}
     \SetKwFunction{ParallelMergeSort}{ParallelMergeSort}
    \SetKwFunction{ParallelMerge}{ParallelMerge}
    \SetKwFunction{GetBloomFilter}{GetBloomFilter}
    \SetKwFunction{SetBloomFilter}{SetBloomFilter}
    \SetKwFunction{Processed}{Processed} 
    \SetKwFunction{Visited}{Visisted} 
    \SetNoFillComment 

      \ForEach{$q_i \in \mathcal{Q}_{\rho}$ in parallel \label{line:parallel_for_outer}}{
         $u^*_i$, $\mathcal{L}_i$ $\gets$ $s$\;
         converged $\gets false$\;
         \While{\textbf{not} converged}{
         $N_i$ $\gets$ \FetchNeighbors{$u^*_i$, $G$} \label{ln:cpu_compute} \coloredtcp{on CPU}\\
         \ShowLn \coloredtcc{CPU sends nbrs $N_i$ to GPU} \label{ln:data_transfer_cpu_gpu}\\
          \ForEach(\coloredtcp{filter nbrs on GPU \label{ln:filter_nbrs}}) {$n \in N_i$ in parallel}{
              \If{\textbf{not} \GetBloomFilter{$i$, $n$}}
              { $N'_i$ $\gets$ $N'_i$ $\cup$ $\{n\}$\;
              \SetBloomFilter{$i$, $n$}}\label{ln:bloom_filter}
              } 

          \For(\coloredtcp{on GPU} \label{ln:for_loop_distcal}){$k \gets $1  to  $|N'_i|$ in parallel }{
             $\mathcal{D}_i[k] \gets$ \CalculateDistanceToQuery{$N'_i[k], q_i$} \label{ln:distnace_calc} \;    
          }
       $ (\mathcal{D}'_i, \mathcal{N}'_i) \gets$ \ParallelMergeSort{$\mathcal{D}_i, N'_i$} \label{ln:sort} \coloredtcp{on GPU} \\
        $\mathcal{L}_i$ $\gets$ \ParallelMerge{$\mathcal{L}_i, \mathcal{D}'_i, \mathcal{N}'_i$} \label{ln:merge} \coloredtcp{on GPU: Merge   $\mathcal{N}'_i$ with worklist $\mathcal{L}_i$ to keep $t$ NNs at most}\\
        $u^*_i \gets\ $ \text{next unvisited nearest node in}\  $\mathcal{L}_i$ \label{ln:getting_u*} \; 
       \ShowLn \coloredtcc{GPU sends node $u^*_i$ to CPU} \label{ln:data_transfer_gpu_cpu}\\
       $converged \gets$ ($\bigwedge_{n \in \mathcal{L}_i}$ \Visited{n} ) 
     } 
       $\mathcal{K}_{i} \gets\ k$\ \text{candidates nearest to}\ $q_i$\ \text{in}\ $\mathcal{L}_i$\;  
       \ShowLn \coloredtcc{$k$-nearest neighbours available on GPU} \label{ln:data_transfer_gpu_cpu1}       
  }
  \end{algorithm2e}

\ours performs greedy search on a proximity graph~\cite{diskann, ggnn, hnsw}.
Algorithm~\ref{alg:querylookup} describes \ours's scheme for batched- query searches on a GPU. 
Since all queries in a batch are independent, 
each query search can run in a separate CUDA \textit{thread-block} (Section~\ref{ss:gpu_arch}) independently (see Line~\ref{line:parallel_for_outer}), resulting in as many thread-blocks as queries.
It utilises the GPU's massive parallelism to maximise throughput, as measured by Queries Per Second (QPS).
\ours{} addresses the challenge of the entire graph index not fitting on GPU efficiently by keeping the index on CPU and selectively retrieving from the CPU the neighbours of the visited nodes (i.e. candidate nodes) during query search iteration (Lines~\ref{ln:cpu_compute} and~\ref{ln:data_transfer_cpu_gpu}).
This strategy ensures effective handling of billion-scale datasets on a single GPU, enhancing the overall performance of large-scale ANNS.

\begin{figure}[t]
  \centering
  \includegraphics[width=0.85\linewidth]{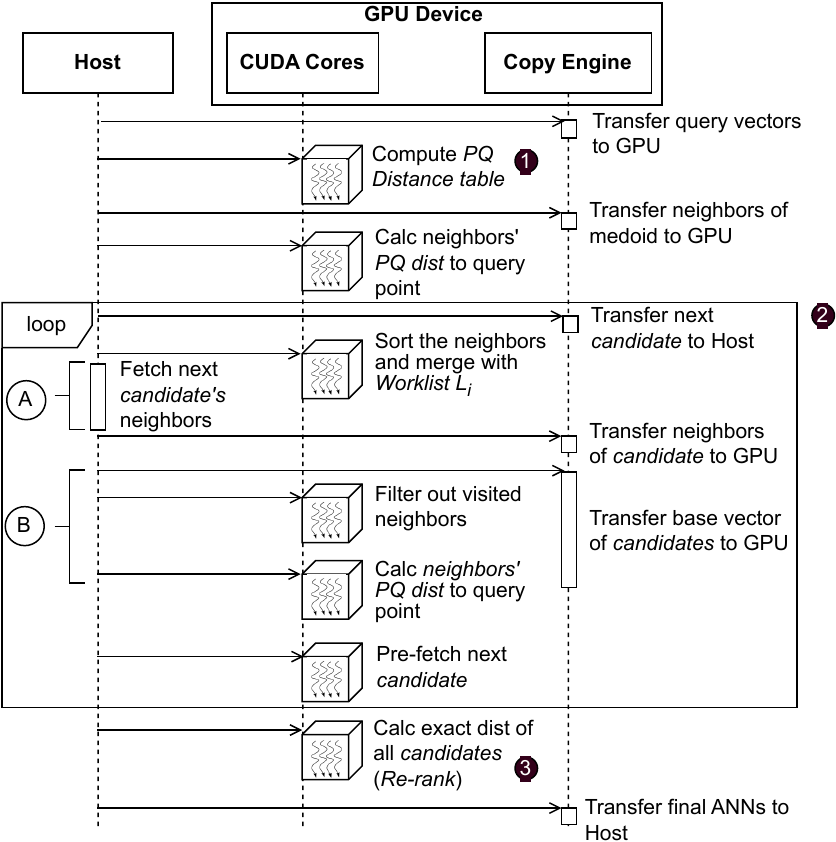}
  \caption{Schema of {\normalfont \ours}.
 }
 \label{fig:arch}
\end{figure}

   \begin{figure*} [ht!]
\subfloat[]{
\label{fig:workflow:a}
 \begin{minipage}[t]{0.3\textwidth}
\vspace{-\topskip} 
\includegraphics[height=\linewidth,width=\linewidth,keepaspectratio]{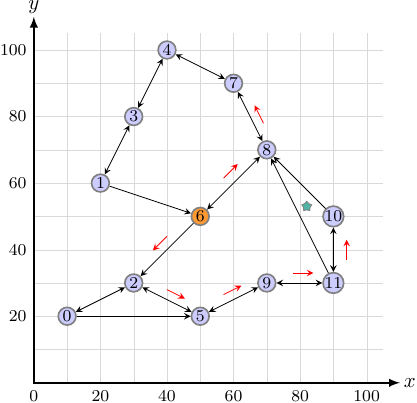}
 \end{minipage}}
 \hfill
 \subfloat[]{
 \label{fig:workflow:b}
  \begin{minipage}[t]{0.17\textwidth}
\vspace{-\topskip} 
\setlength{\tabcolsep}{2pt}
 \scalebox{0.8}{
  \begin{tabular}{r|r|r}
         Node & Cluster & \multicolumn{1}{c}{PQ} \\
         \multicolumn{1}{c|}{ID} & \multicolumn{1}{c|}{ID} & \multicolumn{1}{c}{Distance} \\\hline
         0 &  6& 6273 \\
         1 &  9& 3893\\
         2 &  7& 3233\\
         3 &  8& 3433\\
         4 &  1& 3973\\
         5 &  11& 2113\\
         6 &  0& 1033\\
         7 &  2& 1853\\
         8 &  4& 433\\
         9 &  5& 673\\
        10 &  3& 73\\
        11 &  10& 593\\
     \end{tabular}
     }
 \end{minipage}} 
 \hfill
  \subfloat[]{
  \label{fig:workflow:c}
  \begin{minipage}[t]{0.27\textwidth}
\vspace{-\topskip} 
 \includegraphics[height=\linewidth,width=\linewidth,keepaspectratio]{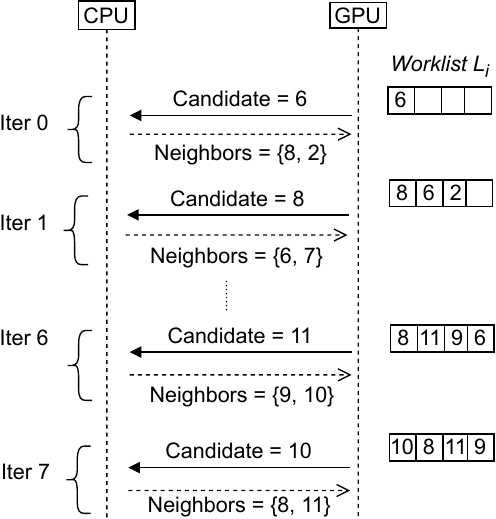}
 \end{minipage}}
 \hfill
\subfloat[]{
\label{fig:workflow:d}
\vspace{2.75\topskip}
  \begin{minipage}[t]{0.2\textwidth}
\vspace{-\topskip} 
\scalebox{0.8}{
\setlength{\tabcolsep}{2pt}
  \begin{tabular}{c|r|r}
         \multirow{2}{*}{Iteration}& \multirow{2}{*}{Candidate} & \multicolumn{1}{l}{Exact} \\
         &  & \multicolumn{1}{l}{Distance}\\\hline
         0 &  6& 1033 \\
         1 &  8& 433\\
         2 &  7& 1853\\
         3 &  2& 3233\\
         4 &  5& 2133\\
         5 &  9& 673\\
         6 &  11& 593\\
         7 &  10& 73\\ 
     \end{tabular}
     }
 \end{minipage}}
  \caption{Example illustrating {\normalfont \ours} workflow for ANNS on \vamana graph.
   The star indicates the query point $q(82,53)$.
   }
   \label{fig:workflow}
\end{figure*}

Figure~\ref{fig:arch} shows the complete workflow of \ours{} and the interaction between CPU and GPU.
CPU maintains the graph and looks up its adjacency list to retrieve the neighbours of a node sent by the GPU.
Similarly, the CPU maintains the complete base dataset as it cannot be fitted on the GPU, and the GPU performs the search using compressed vectors derived from the actual data vectors.
Each step shown under the CUDA Cores section in Figure~\ref{fig:arch} runs in an individual CUDA \textit{thread-block} (Section~\ref{ss:gpu_arch}) independently for each query.
Figure~\ref{fig:workflow} shows the workflow on a toy example, having 12 points in the base dataset, and its prebuilt Vamana graph index is shown in Figure~\ref{fig:workflow:a}.
We use this running example to explain our technique, \ours{}.

\ours{} addresses the challenges discussed in Section~\ref{s:overview} by efficiently utilizing CPU, GPU, and PCIe bus hardware through a phased-execution approach. 
This strategy optimally balances ANNS workload across CPU and GPU while mitigating CPU-GPU data transfer bottlenecks by implementing prefetching techniques for asynchronous data transfer from CPU to GPU during specific phases.
Further, by systematically dividing the ANNS activity into distinct steps,  each with its own characteristics of work distribution and span, we maximise GPU parallelism, in contrast to other approaches~\cite{song,ggnn} that treat the entire search activity as a single block\revchanges{ (i.e. single kernel}).
That is, we design separate GPU kernels \revchanges{(as different phases)} for each of these tasks within the search activity: 
(1) filtering out previously processed neighbours of the current node $u^*_i$ at Line~\ref{ln:filter_nbrs}, 
(2) computing distances of neighbours to the query $q_i$ at Line~\ref{ln:for_loop_distcal},
(3) sorting the neighbours by computed distances at Line~\ref{ln:sort}, and 
(4) merging the closest neighbours into the worklist at Line~\ref{ln:merge}. 
Figure~\ref{fig:workflow:c} shows these steps for each iteration of the query, where the GPU filters and sorts the neighbours by their distance to the query, merges it with $L_i$, and selects the first candidate node (e.g., 8 in iteration 1) to be visited next. 
For each individual kernel, we optimise the thread-block size (per query) to maximise GPU 
occupancy, as a single sub-optimally configured block can significantly reduce performance.
Furthermore, this phased-execution strategy allows us to reduce GPU idle time by eagerly predicting the next candidate node $u^*_i$ immediately after neighbour distance calculation rather than waiting for the previous iteration to finish before communicating $u^*_i$ to the CPU.

In addition to greedy search, \ours employs two additional steps which result in three stages that happen one after the other: \circled{1} the distance table construction, \circled{2} greedy search, and \circled{3} re-ranking, as shown in Figure~\ref{fig:arch}.
Since the dataset cannot fit in GPU memory, it is compressed using the popular Product Quantization~\cite{PQ} technique. 
This technique partitions the $d$-dimensional dataset vector space into $m$ vector subspaces and performs independent $k$-means clustering of the dataset points(nodes) within each subspace, resulting in a centroid vector and ID for each cluster. 
Each dataset point is encoded using the cluster IDs belonging to the point in the respective  $m$ vector subspaces. 
Eventually, the input  $d$-dimensional vector is represented by a shorter $m$-dimensional vector. Figure~\ref{fig:workflow:b} shows the cluster ID associated with each node. Since $m=1$ in the toy example, each 2-dimensional node vector is encoded into a single-byte cluster ID. 
We calculate the distances between centroid vectors and a given query vector within each subspace in a preprocessing step (\circled{1}) and store them in a data structure \emph{PQ Distance Table} (see Figure~\ref{fig:workflow:b}).
During the greedy search stage (\circled{2}), distances between dataset points (in compressed form) and query vectors are calculated by summing the precomputed distances of the cluster IDs across all subspaces (i.e. available in \emph{PQ Distance table}).
Finally, after the greedy search process converges, \ours performs a re-ranking step (\circled{3}) to refine the inaccurate set of $k$ nearest neighbours that might have resulted due to the use of compressed vectors.
The re-ranking step calculates the exact distances (Figure~\ref{fig:workflow:d}) of candidate nodes to the query point using the original data vectors, ensuring precise identification of the final list of $k$ nearest neighbours. (Note: The exact distances in Figure~\ref{fig:workflow:d} and the corresponding PQ distances in Figure~\ref{fig:workflow:b} happen to be identical for the respective nodes as it is a toy example. However, practically, they would be different.)
For the toy example, the sorted final list of candidate nodes, ordered by their distances to the query node, includes {10, 8, 11, 9, 6, 7, 5, 2}, resulting in the selection of the top 2 nearest neighbours, namely, 10 and 8.
\ours performs both reranking and distance table construction on GPU.
Note that, for the re-ranking step, only full vectors of selected nodes are sent to GPU, which are fewer (compared to sending the entire dataset)
and can thus collectively fit into GPU memory for all queries.

In the following subsections, we describe in detail how we parallelise and optimise the mentioned kernels and each step in Algorithm~\ref{alg:querylookup}. 

\subsection{Construction of $PQDistTable$ in Parallel}
\label{ss:pq_dist_construct}





As discussed previously, since the entire dataset cannot fit on the GPU, the search on the GPU uses compressed vectors derived from the actual data vectors.
That is, Algorithm~\ref{alg:querylookup} uses approximate distances calculated using compressed vectors at Line~\ref{ln:distnace_calc}.
For each of the query points in a batch, we compute its squared Euclidean distance to each of the centroids for every subspace produced by the compression. 
We maintain these distances in a lookup data structure, which we call the PQ Distance Table (in short, $PQDistTable$). 
Here, we describe the construction of $PQDistTable$.
We describe later (in Section~\ref{ss:dist_calc}) how we use these precomputed distances to efficiently compute the \textit{asymmetric distance}~\cite{PQ} between a (uncompressed) query point and the compressed data point.

We maintain $PQDistTable$ as a contiguous linear array of size $(\rho \cdot m \cdot 256)$, where $\rho$ is the size of the query batch and $m$ is the number of subspaces with each subspace having 256 centroids. 
The number of centroids is as used in prior works~\cite{faiss,diskann} that use Product Quantization~\cite{PQ}.  
We empirically determine $m = 74$ for our setup, guided by the available GPU global memory (Section~\ref{s:eval} shows the ablation study with varying $m$ values).
For example, in Figure~\ref{fig:workflow:b}, after compression, each input vector (2 bytes) gets compressed to a size of 1 byte using $m = 1$ for the product quantization.
Each thread-block handles one query, resulting in $\rho$ concurrent thread-blocks.
Furthermore, for a given query $q \in \mathbb{R}^d$, the distance of each of the subvectors of the query $q_s \in \mathbb{R}^{d/m}$ 
from each of the 256 centroids of a subspace
can be computed independently of others in parallel by the threads in a thread-block.
Note that, for query $q$, distances of subvectors across $m$ subspaces are computed sequentially within a thread, ensuring an adequate workload per thread and constrained by thread-block size.

\mypara{Work-Span Analysis}
 The work of the algorithm is\\ 
 $O((m \cdot subspace\_size) \cdot 256 \cdot \rho)$.
 Owing to the parallelisation scheme, the span of the algorithm is $O(m \cdot subspace\_size) = O(d)$, where d is the dataset dimension, and $subspace\_size$ is the size of the subspace (i.e., $d/m$).

\subsection{Handling Data Transfer Overheads}
\label{ss:data_transfer}
As outlined in Algorithm~\ref{alg:querylookup}, the CPU transfers the neighbours (Line~\ref{ln:data_transfer_cpu_gpu}) to the search routine executing on the GPU for every candidate transmitted by the GPU (Line~\ref{ln:data_transfer_gpu_cpu}).
The data transfer between the device (GPU) and the host (CPU) is time-consuming in contrast to the GPU's tremendous processing power (the PCIe 4.0 bus connects the GPU and has a maximum transfer bandwidth of only 32 GB/s). 
Therefore, \ours transmits only the bare minimum information required in order to minimise data transfer overhead. 
Specifically, from device to host, the transfer includes a list of final approximate nearest neighbours after the search in  Algorithm~\ref{alg:querylookup} converges and a candidate node for each query in every iteration (Line~\ref{ln:data_transfer_gpu_cpu}), while from host to device, it comprises a list of neighbouring nodes of candidate nodes (Line~\ref{ln:data_transfer_cpu_gpu}) and base vectors/coordinates of the candidates (not shown in the algorithm). 

Further, \ours hides data transfer latency with kernel computations using advanced CUDA features.
To perform data transfers and kernel operations concurrently, CUDA provides asynchronous  \texttt{memcpy} APIs and the idea of streams~\cite{cuda_stream}; \ours makes use of these. 
During the search, neighbours of a given candidate are retrieved using CPU threads for all $\rho$ queries (see Line~\ref{ln:cpu_compute} in Algorithm~\ref{alg:querylookup}).
We leverage the efficient structure of the graph data, allowing sequential memory access of the node's base vector and its neighbourhood list as they are placed next to each other on the host memory.
Hence, immediately after the transfer of the neighbour list to the GPU in each search iteration, we strategically make an asynchronous transfer of the base dataset vectors for future use that are only required during the final re-ranking step on GPU (after the search in Algorithm~\ref{alg:querylookup} converges).
As a result, the kernel execution engine and the copy engine of the GPU are kept occupied to achieve higher throughput.
Thus, in our implementation, we aim to make all the data transfers asynchronous with {\tt cudaMemcpyAsync()} and placing \texttt{cudaStreamSynchronize()} at appropriate places 
to honour the data dependencies.


\revonechanges{
\subsection{Parallel Filtering of Visited Neighbours
}
\label{ss:bloomfilter}

Algorithm~\ref{alg:querylookup} (Line~\ref{ln:cpu_compute}) may encounter the same node multiple times during intermediate iterations of the search. 
Not filtering visited nodes may degrade throughput due to redundant distance computations (Line~\ref{ln:distnace_calc}) and reduce recall by allowing repeated entries in the worklist $\mathcal{L}_i$, potentially displacing more promising nodes and causing premature termination of the search.
Our experiments on the SIFT-1B dataset show that this leads to up to a 10× drop in recall compared to when visited nodes are filtered.
In each search iteration, Line~\ref{ln:cpu_compute} returns up to $R \times \rho$ neighbours to the GPU, where $R$ is the graph’s degree and $\rho$ is the query batch size, necessitating efficient visited checks on this large set to avoid overhead.
Tracking visited nodes on the GPU is challenging, as bitmaps incur high memory overhead for large datasets, and dynamic data structures such as priority queues or hash tables are inefficient on GPU as they underutilise its massive parallelism~\cite{song,ggnn}. 
Therefore, we adopt the well-known Bloom filter for its low memory footprint and suitability for GPU execution.

We use a separate Bloom filter per query and assign one query per thread block to enable fine-grained parallelism (Line~\ref{ln:bloom_filter}).
Parallel accesses to the Bloom filter do not require global synchronisation, as each query uses a separate filter within a thread block. 
In case of a race within a thread block, i.e., if thread $x$ sets an entry for a node and thread $y$ tests a different node that hashes to the same entry before $x$'s write completes, $y$ may see the entry as unset. 
However, this is safe, as it 
prevents incorrect inclusion of the unvisited node of $y$. 
Conversely, in the opposite case, $y$ may see the entry set by $x$ and wrongly mark its own node as visited.
While this can be avoided by synchronizing reads before writes using a barrier like \texttt{\_\_syncthreads()}, we skip synchronisation due to its overheads, and such cases are rare.
}

\subsection{Parallel Neighbour Distance Computation}
\label{ss:dist_calc}



For each query in a batch, we compute its \textit{asymmetric distance} from the current list of neighbours in each iteration (Line~\ref{ln:distnace_calc} in Algorithm~\ref{alg:querylookup}).
This computation for each query is independent, so we assign one query per thread-block.
Further, for a query, its distance to each of the neighbours can also be computed independently of the other neighbours.
Thanks to the $PQDistTable$ we built previously (Section~\ref{ss:pq_dist_construct}), 
we can compute the distance of a query point $q_i$ to a node $n_i$ by summing the \textit{partial} distances of the centroids across all subspaces in $PQDistTable$, with centroid information obtained from $n_i$'s compressed vector.

When computing distances in GPU-based ANNS, summations are commonly conducted using reduction APIs from NVIDIA's CUB~\cite{cub} library, such as those in~\cite{ggnn, song}, typically at a thread-block-level or warp-level. 
However, in our approach, a thread-block of size $t_b$ is subdivided into $g$ groups, each with $g_{size} \coloneqq \frac{t_b}{g}$ threads. 
These groups collaboratively compute the distance of the query from a neighbour, with $g_{size}$ threads summing up the partial distances of $m$ subspaces (with each thread processing a segment of size $\frac{m}{g_{size}}$). 
Each thread sequentially computes the sum of values in a segment using thread-local registers, avoiding synchronisation.
Finally, to efficiently add segment-wise sums using $g_{size}$ threads, we explore two approaches: (i) atomics, employing \textit{atomicAdd()} for the final result; (ii) Sub-warp-level reductions, utilising CUB~\cite{cub} \textit{WarpReduce}.
Empirical tuning with $m = 74$, $t_b = 512$, and $g_{size} = 8$ yields optimal performance in our experimental setup, with the second approach, utilizing sub-warp level reductions, marginally outperforming the first.
Overall, our segmented approach outperforms standard alternatives like CUB \textit{WarpReduce} ($1.2\times$ slower) and \textit{BlockReduce} ($4\times$ slower) for warp-level and thread-block-level reductions.

This kernel constitutes 
a sizeable chunk
of the total run time on billion-size datasets, accounting for approximately 38\% on average. 
The kernel's efficiency is limited by GPU global memory access latency arising from uncoalesced accesses to compressed vectors of various neighbours, a consequence of the irregular structure of the graph.
To mitigate this irregularity,
we explored graph reordering using the well-known Reverse Cuthill Mckee (RCM) algorithm~\cite{cuthill69, george71}.
However, this approach did not yield significant improvements in locality, and so we 
do not perform
graph reordering. 

\mypara{Work-Span Analysis}
The work of the algorithm is\\
$O(R \cdot m \cdot \rho )$, where $R$ is the degree bound of the graph.
Owing to the parallelisation scheme, the span of the algorithm is $O(log~m)$.

\subsection{Prefetching Candidate Nodes}
\label{s:prefetching}
As described in Algorithm~\ref{alg:querylookup}, 
candidate node  $u^*_i$ of each query $q_i$ will be communicated to the CPU (Line~\ref{ln:data_transfer_gpu_cpu}) only after the GPU has updated the worklist $\mathcal{L}_i$. 
After that, while the CPU fetches the neighbours (Line~\ref{ln:cpu_compute}) and communicates them to the GPU (Line~\ref{ln:data_transfer_cpu_gpu}), the GPU remains idle, waiting to receive this information. 
To avoid this delay in starting the current iteration, GPU must be supplied with the neighbour IDs as soon as needed. 
To address this, instead of sending the candidate nodes at the end of the previous iteration to CPU, 
we perform an optimisation to  
\textit{eagerly} identify the next candidate nodes immediately after the neighbour distance calculation (i.e., before even sorting the neighbour list by distance and updating worklist $\mathcal{L}_i$).

Our optimisation calculates the candidate node $u^*_i$ eagerly by selecting the nearest neighbour in the new neighbour list $N'_i$ and the first unvisited node in worklist $\mathcal{L}_i$ (sorted by distance to query), then choosing the best between the two. 
The eager selection occurs just before Line~\ref{ln:sort} in Algorithm~\ref{alg:querylookup}.
Upon dispatching the eagerly selected candidates to the CPU, the GPU continues executing the remaining tasks of the iteration (sorting the new neighbour list and updating the worklist). Concurrently, the CPU fetches neighbour IDs.
That is, Line~\ref{ln:cpu_compute}
executes on CPU concurrently with tasks at Line~\ref{ln:sort} and~\ref{ln:merge} on GPU.
By the completion of the GPU's remaining tasks, the CPU has already communicated the neighbour IDs. 
This strategy eliminates GPU idle time, resulting in an increase in throughput and experimentally found to be $8-12\%$.

\subsection{Parallel Merge Sort}
\label{ss:parallelmergesort}
To add new vertices closer to $q_i$ than those in $\mathcal{L}_i$ in each iteration of Algorithm~\ref{alg:querylookup}, we sort the neighbours $N'_i$ in the increasing order of their 
asymmetric distances to $q_i$ and attempt to merge the eligible ones with $\mathcal{L}_i$.
We sort the neighbours using parallel bottom-up merge sort.
Conventionally, one worker thread merges two lists (in the conquer phase of the algorithm).
Thus, as the algorithm progresses, parallelism decreases since there are fewer lists to merge, and consequently, the amount of work per thread increases as the size of the lists to merge grows. 
As most GPU threads remain idle until the sort is complete and the work per thread is high, this scheme is not suitable for GPU processing. 
We mitigate these issues by merging lists in parallel, using a parallel merge routine described in the following subsection.

For our use case, we need to sort small lists, typically having up to 64 nodes. 
We assign one thread-block to a query. 
Furthermore, we assign one thread per neighbour in the list to be sorted by setting the thread-block size to the maximum number of neighbours of a node in the graph.
We start with sorted lists that have one element each and double their size at each step through the parallel merge routine (Section~\ref{ss:parmerge}).
As the lists are small, we can keep them in GPU shared memory 
throughout the sorting algorithm.

\mypara{Work-Span Analysis}
The work of merging two lists of size $t$ using the parallel merge routine is $O(t \cdot log(t))$ (Section~\ref{ss:parmerge}).
The total work $T(n)$ for an array of $n$ numbers follows the recurrence:
$T(n) = 2 \cdot T(n/2) + n \cdot log(n)$, 
which gives $T(n) = O(n \cdot log^2(n))$.
Thus, the work of the algorithm is $O(R \cdot log^2(R) \cdot \rho)$. 
Owing to the parallelisation scheme, the span of the algorithm is $O(log^2(R))$.

\subsection{Parallel Merge}\label{ss:parmerge}

\begin{figure}
  \centering
  \includegraphics[width=0.9\linewidth]{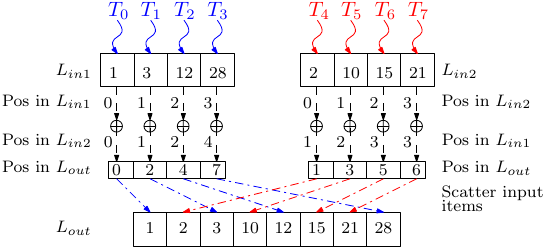}
   \caption{Parallel List Merge. $L_{in1}$ and $L_{in2}$ are merged to create the merged list $L_{out}$.}
   \label{fig:parmerge}
  \end{figure}
The parallel merge routine that we detail in this section is employed in the parallel merge sort, as explained in the preceding section. 
Additionally, it is utilised to merge the sorted neighbour list $\mathcal{N}'_i$ with the worklist $\mathcal{L}_i$ at Line~\ref{ln:merge} in Algorithm~\ref{alg:querylookup}. 

      



Figure~\ref{fig:parmerge} shows our parallel list merge algorithm on two lists of size four each.
Consider two sorted lists $L_{in1}$ and $L_{in2}$ having $m$ and $n$ items respectively. Suppose that an item $e$ in $L_{in1}$ is at position $p_1 < m$, and if inserted in $L_{in2}$ while maintaining sorted order, it will be placed at position $p_2 < n$. Then, the position $pos_{Lout}$ of the element in the merged list $L_{out}$ is determined by $pos_{Lout} \gets p_1 + p_2$.
We assign one thread to each element. 
From the thread ID assigned to the item, we can compute the position of the item in its original list.
We use binary search to determine where the item should appear in the other list.
For example, in Figure~\ref{fig:parmerge}, the item 28 in $L_{in1}$ is at index 3, determined by the thread ID. 
In $L_{in2}$, its index is 4, found through binary search. Therefore, the index of 28 in the merged list is 7 (i.e., 3 + 4).
Each thread, assigned to an item, concurrently performs a binary search for that item in the other list. 
We keep the lists in the GPU's shared memory to minimise search latency. 
Thus, computing the position of each element in the merged list takes $O(\log(m) + \log(n))$ time. 
Lastly, elements are scattered to their new positions in the merged list; in the example, 28 is placed at index 7 in the merged list $L_{out}$. 
This step is also performed in parallel, as each thread writes its item to its unique position in the merged list.

  \mypara{Work-Span Analysis}
The work of the algorithm to merge two lists each of size $\ell$ is $O(\ell \cdot log(\ell))$. 
Owing to the parallelisation scheme, the span of the algorithm is
$O(log(\ell))$.

\subsection{Re-ranking}
\label{s:re-ranking}

As discussed, we use approximate distances (computed using PQ distances) throughout the search in Algorithm~\ref{alg:querylookup}.
Hence, we employ a final re-ranking step~\cite{Wieschollek_2016_CVPR} after the search converges to enhance the overall recall.  

We implement the re-ranking step as a separate kernel with optimised thread-block sizes. 
Re-ranking involves computing the exact (i.e. Euclidean) distance for each query vector with its respective candidate nodes, followed by sorting these candidates by their distance to the query vector and reporting the \textit{top-$k$} candidates as nearest neighbours.
For a query, its distance from each candidate node is computed in parallel.
To sort the candidates according to their exact distance, we use the parallel merge procedure described in Section~\ref{ss:parallelmergesort}.
Thanks to the asynchronous data transfers (Section~\ref{ss:data_transfer}), the candidates' base dataset vectors are already available on GPU when this kernel begins. 
From our experiments, we observed that re-ranking improved the recall by 10-15\% for the datasets under consideration.

\mypara{Work-Span Analysis}
The work of the re-ranking step is \\ $O((d \cdot |\mathcal{C}|   + |\mathcal{C}| \cdot log^2(|\mathcal{C}|))\cdot \rho)$, where $|\mathcal{C}|$ is the maximum number of candidate nodes for a query and $d$ is vector dimension.
  Owing to the parallelisation scheme, the span of re-ranking is $O(d + log^2(|\mathcal{C}|))$. 
\section{Implementation Variants}
\label{s:impl}

We refer to the implementation described so far in Sections~\ref{s:overview} and~\ref{s:kernels} as \ours. 
In addition, we provide versions of the implementation specifically optimised for medium-sized datasets (having up to 100-million points) that can fit in the GPU memory along with their respective indexes.
We discuss these versions below.

\subsection{In-memory Version}
\label{s:in-memory}

\begin{figure}[t]
  \centering
  \includegraphics[width=0.8\linewidth]{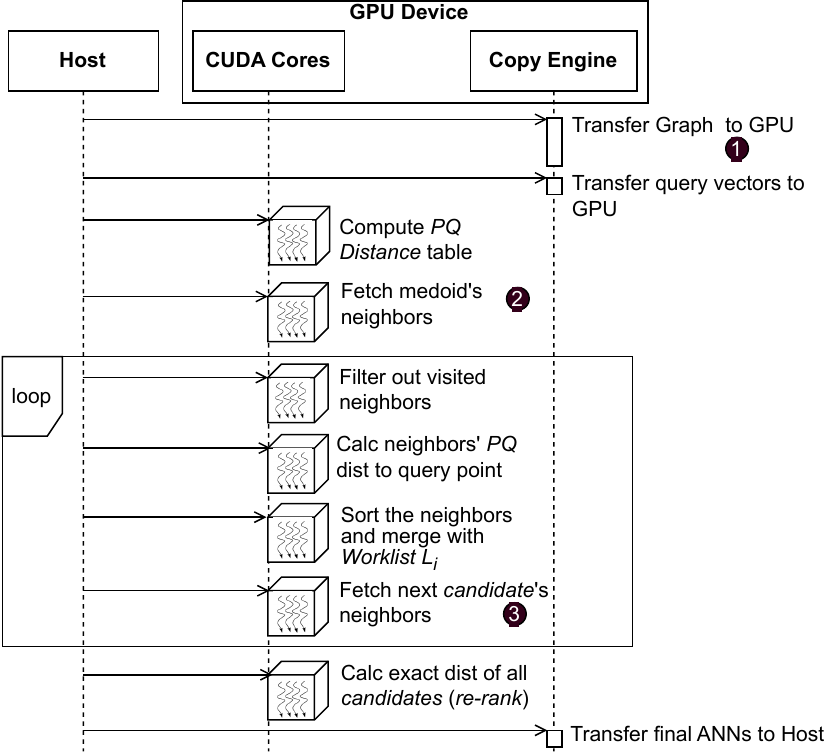}
  \caption{Schema of {\normalfont \banginmemory} Variant.}
 \label{fig:inmemory}
\end{figure}

Since the CPU-GPU interconnect stays constantly busy when we store the graph on the CPU, as in \ours, careful task management on the GPU is needed to avoid idle time due to CPU-GPU interdependency.
However, following the same approach is unnecessary for smaller graphs that can fit on a GPU. 
Keeping the entire graph on the GPU, rather than the CPU, 
eliminates the memory-intensive process of the CPU fetching neighbours from its memory for GPU-supplied candidates, 
resulting in enhanced throughput. 
We refer to this version of the implementation as \banginmemory.
The schematic block diagram of this version is shown in Figure~\ref{fig:inmemory}. 
There are two main differences from Figure~\ref{fig:arch}: 
(1) in step \circled{1}, we transfer the entire graph from CPU to GPU before the search begins
(2) steps \circled{2} and  \circled{3}  fetch neighbours on the GPU locally through accesses to global memory instead of relying on the CPU.

\subsection{Exact-distance Version}
\label{s:exactDistance}

This variant further optimises the \banginmemory version. Instead of using PQ distances within the greedy search loop and subsequently compensating the inaccuracies (resulting due to the use of PQ distances) by the re-ranking step, we directly use exact (Euclidean) distance calculations instead of PQ distances. 
We refer to this version as \bangexactdist.
The main differences from Figure~\ref{fig:inmemory} are: 
(1) absence of PQ Distance Table computation,
(2) exact distance computation (instead of PQ distance summation),
(3) absence of re-ranking step. 
\section{Experimental Setup}
\label{s:eval}


\subsection{Machine Configuration\label{subsec:exptsetup}}


\mypara{Hardware}
We conduct our experiments on a machine with 2 x Intel Xeon Gold 6326 CPU (with 32 cores), clock-speed 2.90GHz, 
and  660 GiB DDR4 memory. 
\revchanges{(For the Text-to-Image-1B dataset, we used 1133 GiB  due to its large size.)}
It houses a 1.41 GHz NVIDIA Ampere A100 GPU with 80 GB global memory and a peak bandwidth of 2039 GB/s.
The GPU is connected to the host via a PCIe Gen 4.0 bus having a peak transfer bandwidth of 32 GB/s. 
We use a single GPU for all the experiments.


\mypara{Software}
The machine runs Ubuntu 22.04.01 (64-bit). 
We use \textit{g++} version 11.3 with \texttt{-O3}, \texttt{-std=c++11} and \texttt{-fopenmp} flags to compile the
C++ code.
We use OpenMP for parallelising the 
C++ code.
The GPU CUDA code is compiled using \textit{nvcc} version 11.8  with the \texttt{-O3} flag. 
The source code of \ours is publicly available at \textit{\url{https://github.com/karthik86248/BANG-Billion-Scale-ANN}}. 

\subsection{Datasets \label{subsec:exptdataset}}
\begin{table}[t]
\setlength{\tabcolsep}{3pt}
\caption{Real-world datasets in our test-suite. (CV Size stands for Compressed Vectors size)}
\small
 \resizebox{\columnwidth}{!}{
\begin{tabular}{l|l|r|r|r|r|r}
& & \multicolumn{1}{c|}{\textbf{Number of}} & & \textbf{Vector} & \multicolumn{1}{c|} 
 {\textbf{Graph}} & \multicolumn{1}{c} {\textbf{CV}} \\
 \multicolumn{1}{c|}{\textbf{Dataset}} & \textbf{Type} & \multicolumn{1}{c|}{\textbf{Data Points}}
 & \textbf{$d$}
 & \textbf{Size\tiny(GB)} & \textbf{Size\tiny(GB)} & \textbf{Size\tiny(GB)} \\
 \hline
 SIFT-1B \cite{sift1b}& uint8& 1000,000,000& 128& 128.0 & 260.0& 74.0\\ 
 Deep-1B \cite{deep1b}& float& 1000,000,000 & 96& 384.0 & 260.0& 74.0\\ 
 MSSPACEV-1B \cite{spacev1b} 
 & int8& 1000,000,000& 100& 100.0 & 260.0& 74.0\\
 \revchanges{Text-to-Image-1B \cite{Text2ImageDataset} }
 & \revchanges{float}& \revchanges{1000,000,000}& \revchanges{200}& \revchanges{800.0} & \revchanges{260.0}& \revchanges{74.0} \\
 SIFT-100M & uint8& 100,000,000&  128& 12.8 & 26.0 & 7.4\\
 Deep-100M & float& 100,000,000& 96& 38.4 & 26.0 & 7.4\\ 
 MSSPACEV-100M & int8& 100,000,000&  100& 10.0 & 26.0 & 7.4\\
 \revchanges{Text-to-Image-100M} & \revchanges{float}& \revchanges{100,000,000}&  \revchanges{200}& \revchanges{80.0} & \revchanges{26.0} & \revchanges{7.4}\\
\end{tabular}
 }
\label{dataset}
\end{table}

We present experiments on four popular real-world datasets 
that originate from a diverse set of applications.
 We summarise the characteristics of 
 these datasets along with the sizes of the graphs generated using Vamana and compressed vectors in Table~\ref{dataset}.
The \textit{Deep dataset}~\cite{deep1b} is a collection of one billion image embeddings compressed to 96 dimensions. 
The \textit{SIFT dataset}~\cite{sift1b} dataset represents the 128-dimensional SIFT (Scale-Invariant Feature Transform) descriptors of one billion images. 
The \textit{Microsoft SPACEV (MSSPACEV) dataset}~\cite{spacev1b} encodes web documents and web queries sourced from Bing using the Microsoft SpaceV Superior Model. 
This dataset contains more than one billion points, and to be consistent with other billion-scale datasets in our test suite, we pick the first billion points in the dataset.
\revchanges{
The Text-to-Image (T2I) dataset, derived from the Yandex~\cite{Text2ImageDataset} visual search engine, comprises one billion 200-dimensional image embeddings for indexing. 
It focuses on a cross-domain setting, where a user provides a textual query, and the search engine retrieves the most relevant images. 
This scenario introduces the possibility of differing distribution patterns between the dataset and the query vectors, also referred to as an Out-Of-Distribution (OOD) dataset, which makes it challenging to handle.
Henceforth, for brevity, we refer to datasets containing one billion data points as \textit{1B datasets} and those with 100 million data points as \textit{100M datasets}. 
The 100M versions of the four datasets are generated by extracting the first hundred million points from the respective 1B datasets. 
We use a query batch size of 10,000 unless otherwise stated, with all queries executed concurrently in a single batch, following prior works~\cite{song,ggnn}. 
For datasets that provide more than 10,000 queries, we select a subset consisting of 10,000 queries. 
}

\subsection{Algorithm Parameters: {\normalfont \ours} \label{subsec:bangparam}}
\mypara{Graph Construction} We run \ours search algorithm on Vamana graph, built using \diskann~\cite{diskann}. 
We specify $R =64$ (maximum degree bound), $L=200$ (size of worklist used during build) and $\sigma$=1.2 (pruning parameter), following the recommendations of the original source.
The rightmost column in Table~\ref{dataset} shows the size of the resulting graph for each dataset. 

\revonechanges{
\mypara{Bloom filter} 
We use a Bloom filter per query,  of size $\sim 400$ kB.
Our Bloom filter uses an array of $z$ \textit{bools}, 
where $z$ is determined using an estimate of the number of processed nodes for a query, a small tolerable false-positive rate, and the number of hash functions used. 
We use two FNV-1a hash functions~\cite{fnv}, which are lightweight, non-cryptographic hash functions often used to implement Bloom filters.
}

\mypara{Compression} During graph construction, we configure the parameters for PQ compression such that the generated compressed vectors fit entirely in the GPU global memory. 
For our setup, we empirically determined the largest value of $m$ to be 74. 
We use $m = 74$ across all datasets for all our experiments unless otherwise stated. 
Note that as a consequence, the compression ratio could vary (although $m$ remains the same) for different datasets depending upon their dimensionality. For example, for the 128-dimension SIFT-1B dataset, the compression ratio is $74/128 = 0.57$, while for the 384-dimension Deep-1B dataset, the compression ratio is $74/384 = 0.19$.
The rightmost column in Table~\ref{dataset} shows the total size of PQ compressed vectors for each dataset.

\mypara{Search} 
The search starts from the \textit{medoid}, a 
node pre-determined during graph construction.
We use the widely used $k\mbox{-}recall@k$~\cite{diskann} recall metric. 
\revchanges{The goal of ANNS is to efficiently retrieve, for a query point $q$, a set $\tilde S$ of $k$ candidate points from the dataset in order to maximise $k$-$recall@k \coloneq \frac{|S \cap \tilde S|}{k}$, where $S$ is the ground truth of the $k$ nearest neighbours of $q$ in the dataset. 
}
We measure the throughput using 
\textit{Queries Per Second} or \textit{QPS} which is \revchanges{
(\textit{number of queries processed / search time}), where search time is the time taken to process all queries.
} 
The search parameter $t$ in Algorithm~\ref{alg:querylookup} represents the size of the worklist $\mathcal{L}$.
The recall increases with $t$. 
The lower bound for $t$ is $k$, and the maximum value of $t$ is empirically set to 152 for our experiments.
The Bloom filter (Section~\ref{ss:bloomfilter}) is created to hold 399,887 entries \revchanges{per query}, which we determined empirically. 
We tune the Bloom filter size to lower values in order to generate lower recall values (below the recall values achieved by using $t=k$).  
To increase statistical significance, we report the average throughput 
over five independent runs. 

\subsection{Baselines \label{subsec:exptcomparison}}
\revchanges{We primarily evaluate \ours against the state-of-the-art open-source ANNS methods GGNN~\cite{ggnn} and FAISS~\cite{faiss} that support billion-scale data on a single GPU. 
While our focus is on billion-scale datasets, we include comparisons on 100M datasets for completeness, including comparison with CAGRA~\cite{cagra}, the leading GPU-based ANNS method for 100M points, despite its lack of support for billion-scale data. }\\
\mypara{GGNN~\cite{ggnn}} \revonechanges{ GGNN provides configurable graph construction parameters
\textit{viz.} number of out-going edges ($k$), number of disjoint points selected for subgraphs ($s$), number of levels in the hierarchical graph ($l$), slack factor ($\tau_b$) and number of refinement iterations ($r$).
In the original work, eight GPUs were used for billion-scale datasets, but we configured GGNN to run with a single GPU.
We use the recommended build parameters from the original source code repository~\cite{GGNNGitHub} for most datasets; for others, we follow authors' private email suggestions for optimal performance on our setup.
We configure 1B dataset as \{$k$:20, $s$:32, $l$:4, $\tau_b$:0.6, $r$:2, 16 shards\} and 100M dataset as \{$k$:24, $s$:32, $l$:4, $\tau_b$:0.5, $r$:2, 2 shards\}. 
}

\begin{figure*}[t]
\centering
  \includegraphics[width=0.4\linewidth]{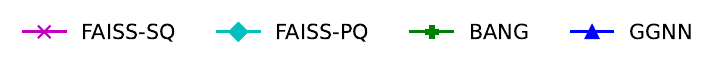}\\
\centering 
 \begin{subfigure}{0.24\linewidth}
    \centering
    \includegraphics[width=\linewidth]{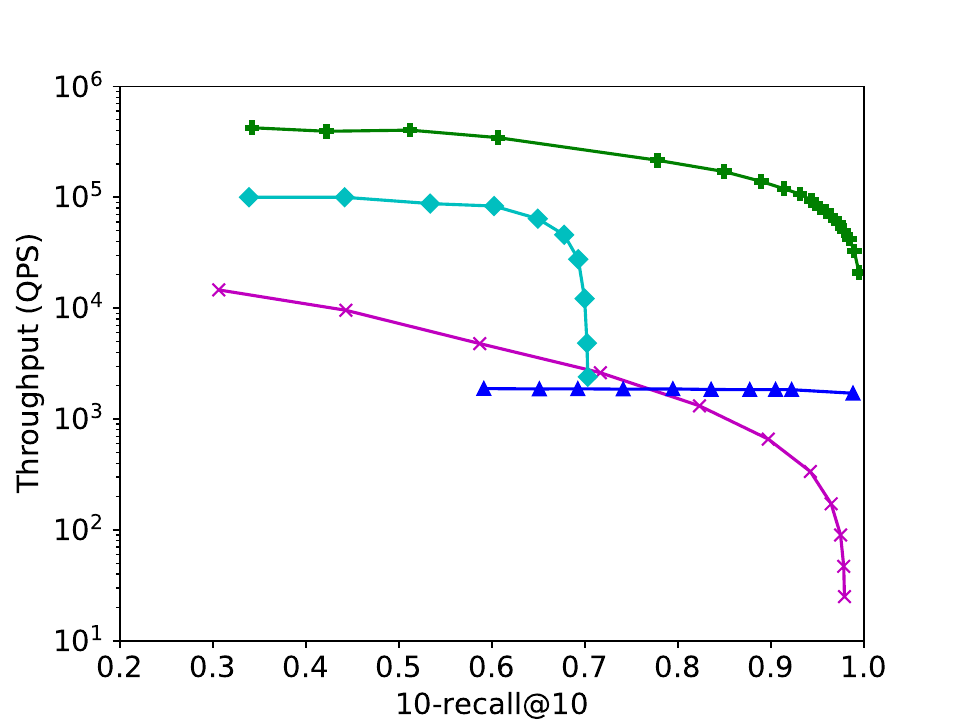} 
    \caption{SIFT-1B}
    \label{fig:sift1b}
 \end{subfigure}%
 \hfill
\begin{subfigure}{0.24\linewidth}
    \centering
    \includegraphics[width=\linewidth]{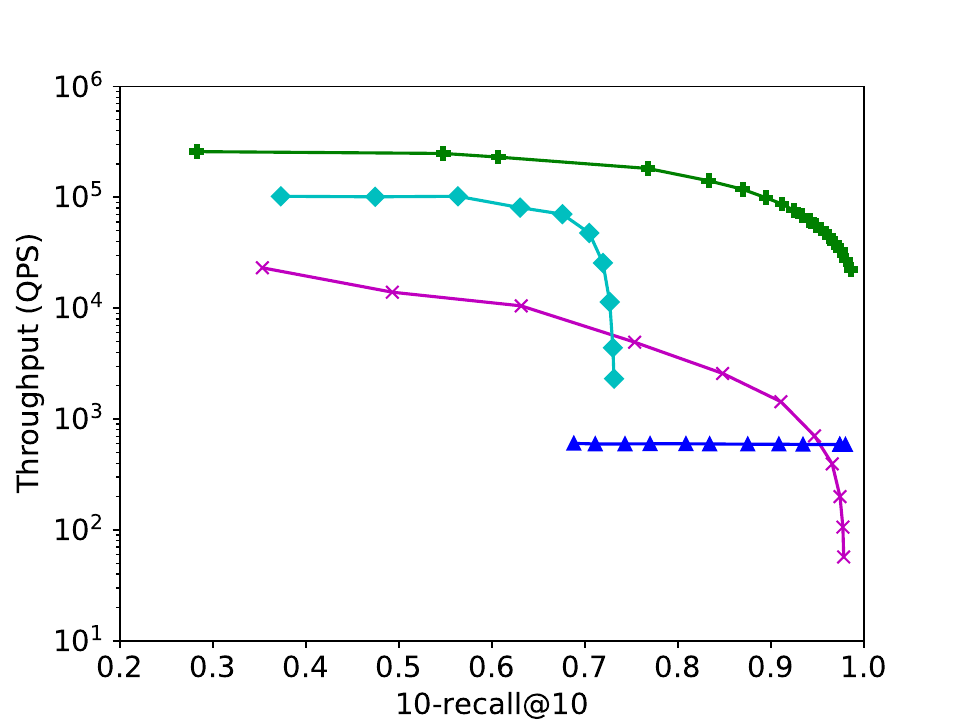} 
    \caption{Deep-1B} 
    \label{fig:deep1b}
 \end{subfigure}%
 \hfill
  \begin{subfigure}{0.24\linewidth}
    \centering
    \includegraphics[width=\linewidth]{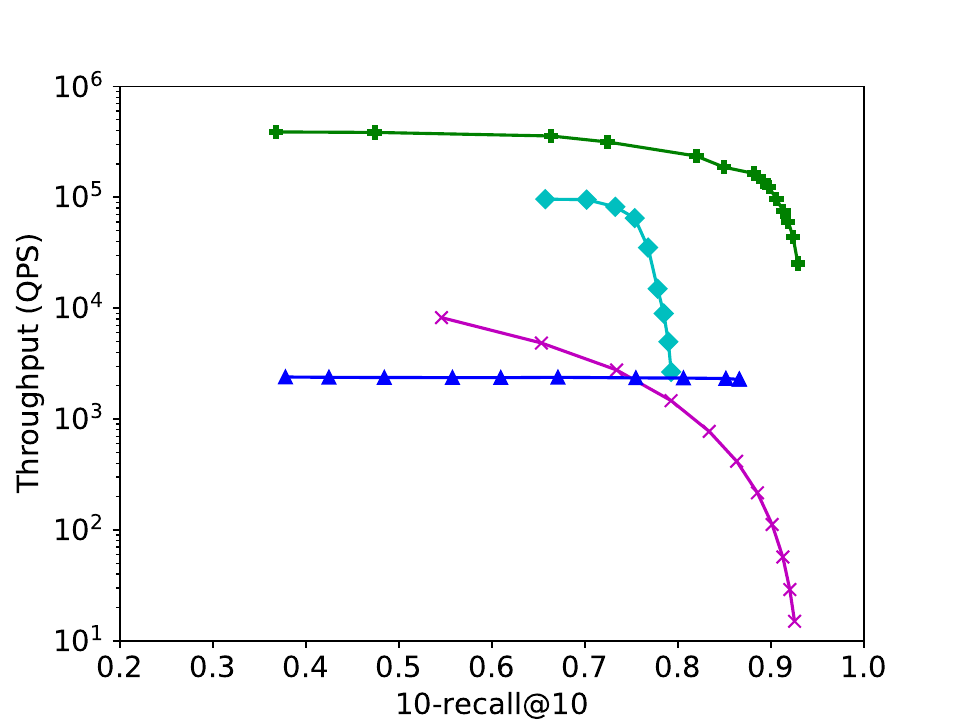} 
    \caption{MSSPACEV-1B}
    \label{fig:spacev1b}
 \end{subfigure}%
 \hfill
\begin{subfigure}{0.24\linewidth}
    \centering
    \includegraphics[width=\linewidth]{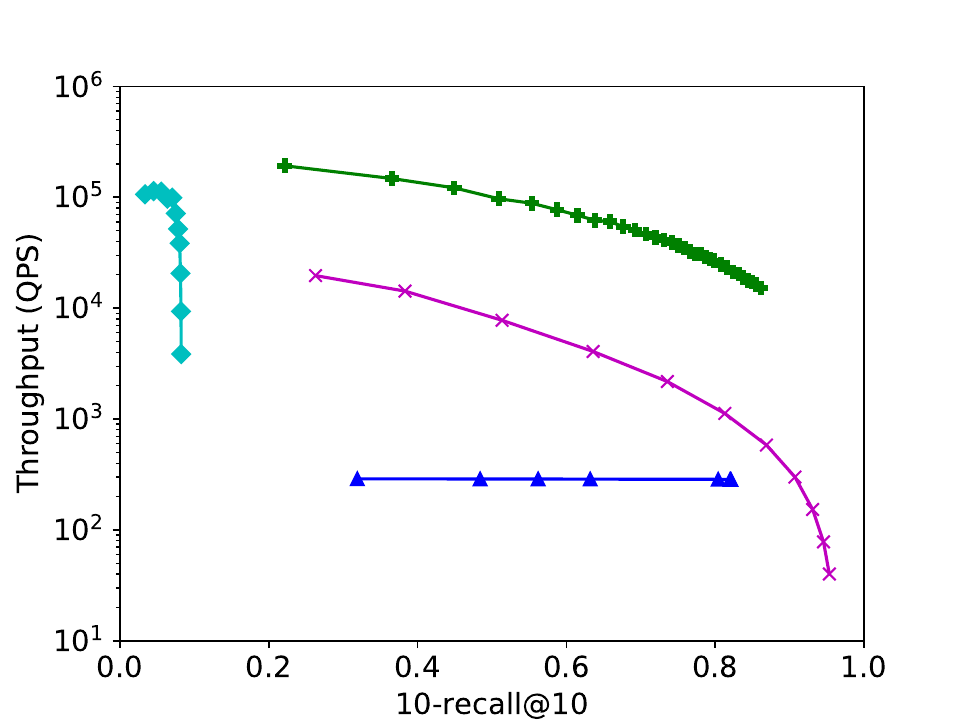} 
    \caption{\revchanges{Text-to-Image-1B}} 
    \label{fig:text2image1b}
 \end{subfigure}
  \hfill
    \caption{\revchanges{Throughput (y-axis) v/s 10-recall@10 for 1B 
    datasets. y-axis is in log-scale.}}
    \label{fig:1b_datasets}
\end{figure*}

\begin{figure*}[t]
\centering
  \includegraphics[width=0.4\linewidth]{figures/legend1b.pdf}\\
\centering 
 \begin{subfigure}{0.24\linewidth}
    \centering
    \includegraphics[width=\linewidth]{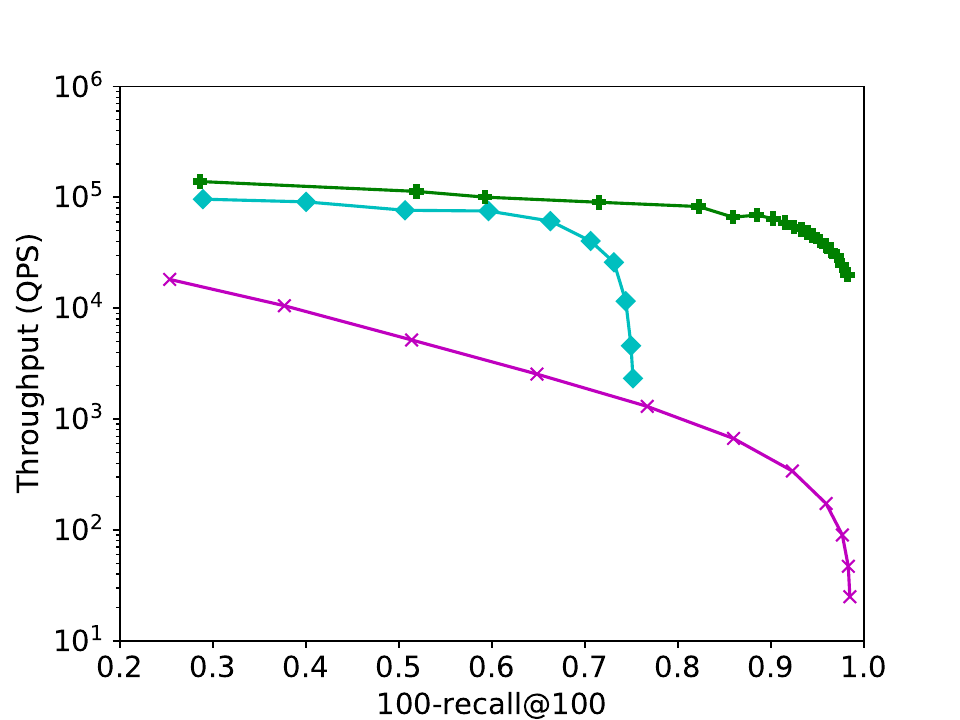} 
    \caption{\revchanges{SIFT-1B}}
    \label{fig:sift1b_r100}
 \end{subfigure}%
 \hfill
\begin{subfigure}{0.24\linewidth}
    \centering
    \includegraphics[width=\linewidth]{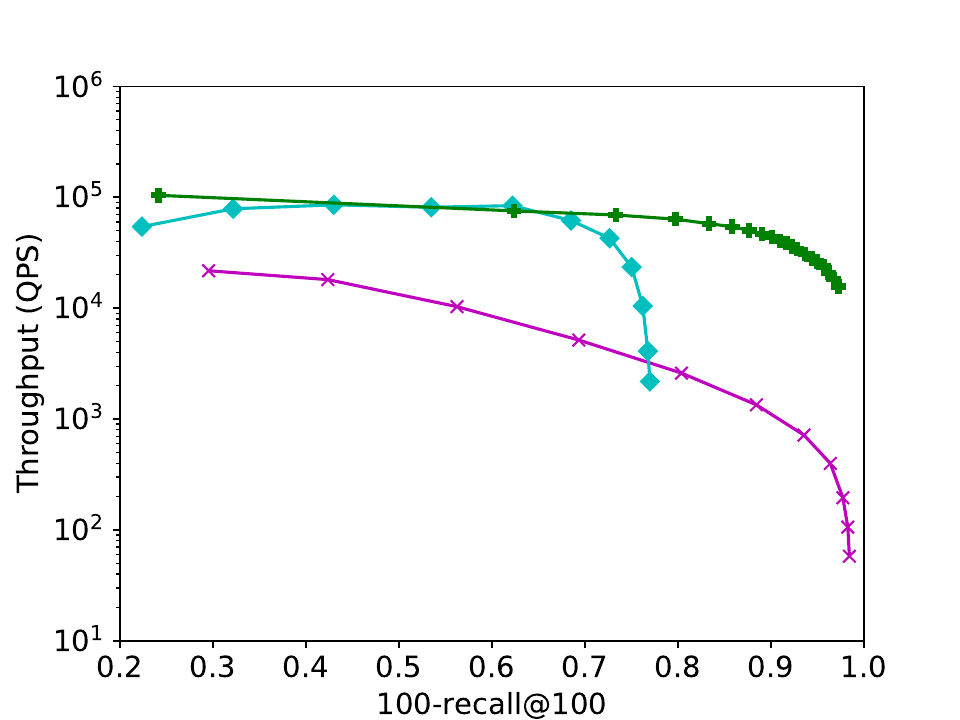} 
    \caption{\revchanges{Deep-1B}} 
    \label{fig:deep1b_r100}
 \end{subfigure}%
 \hfill
  \begin{subfigure}{0.24\linewidth}
    \centering
    \includegraphics[width=\linewidth]{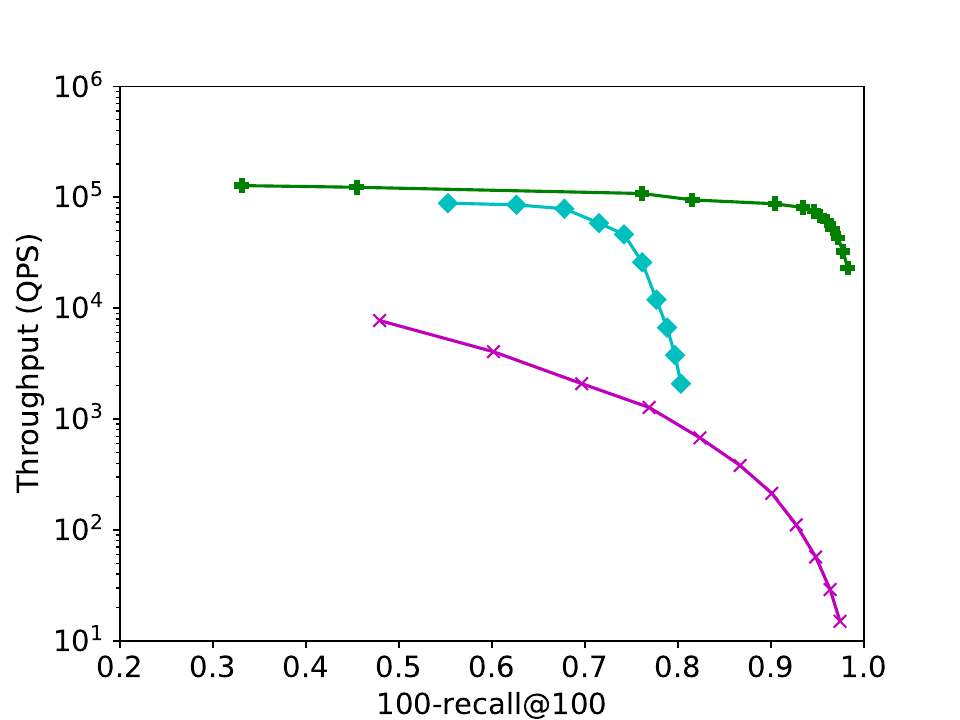}
    \caption{\revchanges{MSSPACEV-1B}}
    \label{fig:spacev1b_r100}
 \end{subfigure}%
 \hfill
\begin{subfigure}{0.24\linewidth}
    \centering
    \includegraphics[width=\linewidth]{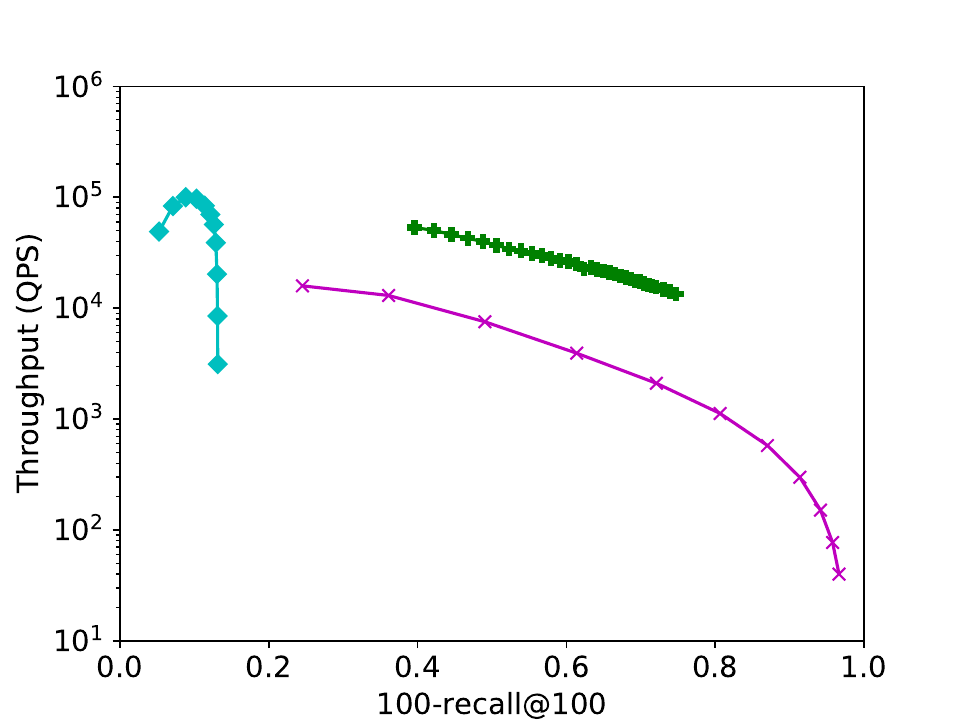} 
    \caption{\revchanges{Text-to-Image-1B}}
    \label{fig:text2image1b_r100}
 \end{subfigure}
    \caption{\revchanges{Throughput (y-axis) v/s 100-recall@100 for 1B 
    datasets. y-axis is in log-scale.}}
    \label{fig:1b_datasets_r100}
\end{figure*}

\begin{figure*}[t]
\centering
    \includegraphics[width=0.9\linewidth]{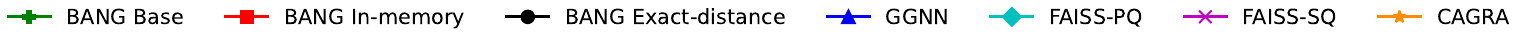}\\
\centering
\begin{subfigure}{0.24\linewidth}
    \centering
    \includegraphics[width=\linewidth]{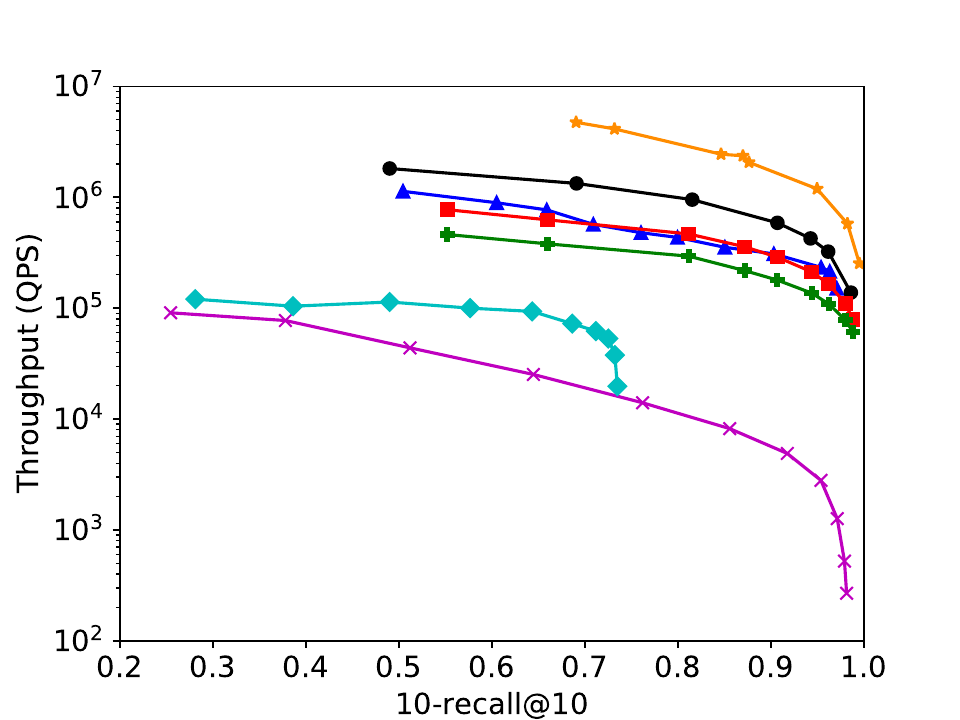}
    \caption{\revchanges{SIFT-100M}}
    \label{fig:sift100m} 
 \end{subfigure}%
\hfill   
\begin{subfigure}{0.24\linewidth}
    \centering
    \includegraphics[width=\linewidth]{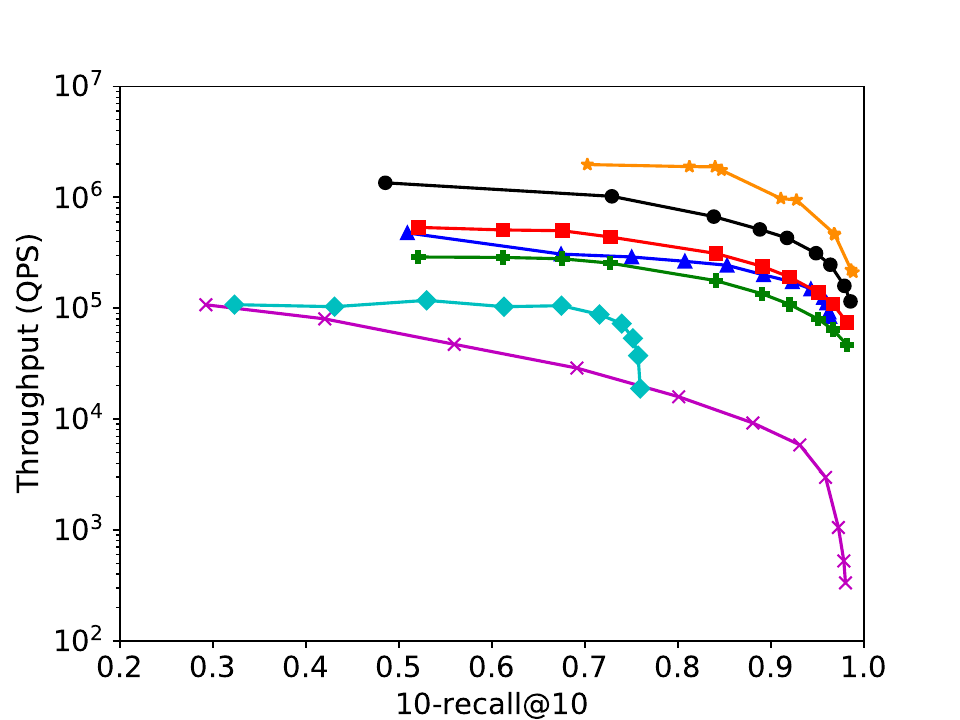}
    \caption{\revchanges{Deep-100M}}
 \label{fig:deep100m}
\end{subfigure}
 \hfill
\begin{subfigure}{0.24\linewidth}
    \centering
    \includegraphics[width=\linewidth]{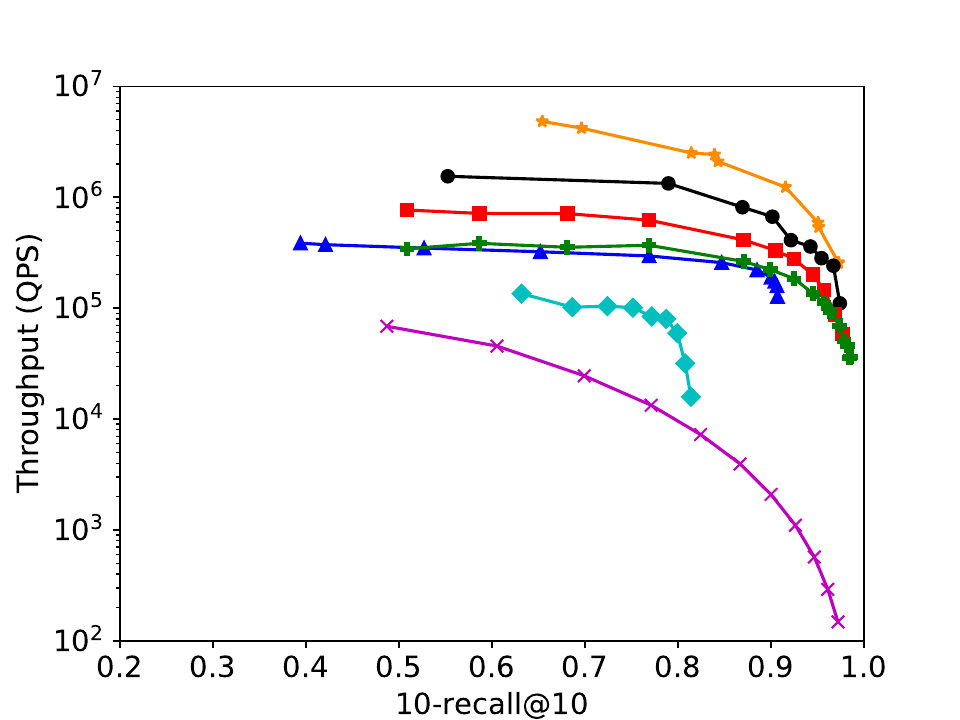}
    \caption{\revchanges{MSSPACEV-100M}}
    \label{fig:spacev100m} 
\end{subfigure}%
 \hfill
\begin{subfigure}{0.24\linewidth}
    \centering
    \includegraphics[width=\linewidth]{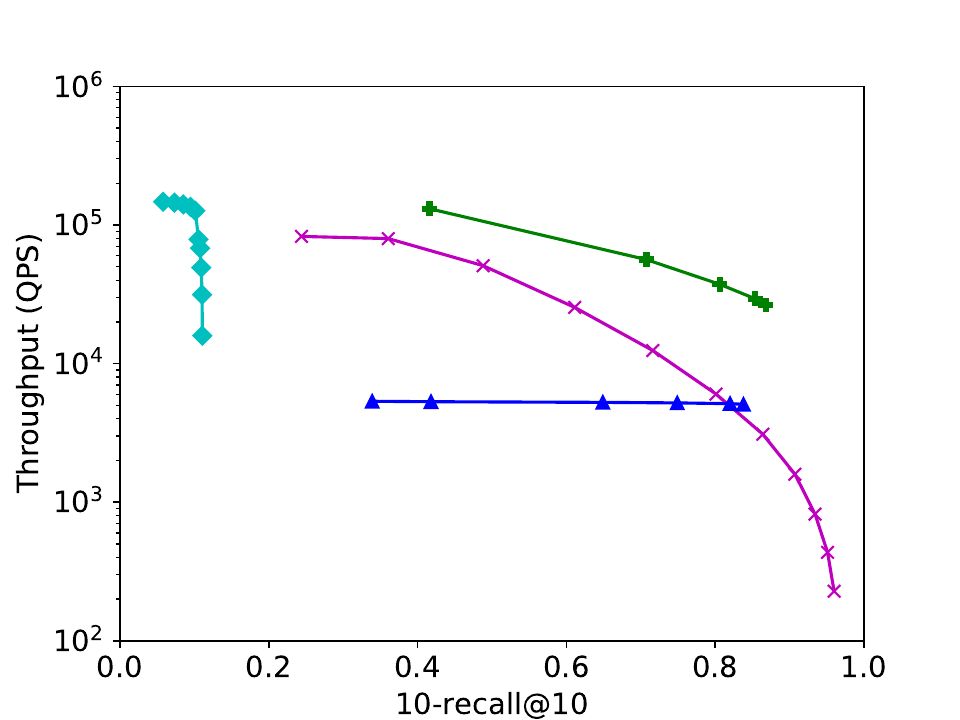} 
    \caption{\revchanges{Text-to-Image-100M}}
    \label{fig:text2image100m}
\end{subfigure}%
    \caption{\revchanges{Throughput (y-axis) v/s 10-recall@10 for 100M datasets. y-axis is in log-scale. 
    }}
    \label{fig:deep100m-sift100m}
\end{figure*} 


\revonechanges{
\mypara{FAISS~\cite{faiss}} 
FAISS provides users with configuration options for quantizing the base dataset using either Scalar Quantization (SQ) or Product Quantization (PQ); following the FAISS Wiki guidelines~\cite{FAISSWiKi}, we experiment with different IVF index types and PQ/SQ transforms to optimise recall and throughput.
For PQ, we configured the index string as: \textit{OPQ32\_128}, \textit{IVF262144}, \textit{PQ32}. 
The first field specifies the vector transformation used for preprocessing, the second denotes the IVF index type, and the third indicates the quantization scheme applied.
For SQ, we configured the index string as:  \textit{IVF65536, SQ8} as indicated in \texttt{Billion-Scale Approximate Nearest Neighbor Search Challenge}~\cite{track3baseline}. 
Note that SQ indexes are less space-efficient than PQ but provide better recalls. 
We performed all experiments on a single GPU.
}

\revchanges{
\mypara{CAGRA~\cite{cagra}} 
CAGRA is a state-of-the-art ANNS algorithm in the Nvidia cuVS~\cite{RAFTANNBenchmarks}, a high-performance, GPU-accelerated library for efficient vector search.
We deployed CAGRA using the RAFT's ANN  benchmark collection.
We configured CAGRA with the NN-Descent~\cite{NN-Descent} algorithm and set the kNN graph degree to 32. 
For optimal search results, we vary the internal priority queue size (\textit{itopk} parameter) from 32 to 512. 
}

\section{Results\label{s:results}}


In this section, we comprehensively compare the performance and solution quality of \ours with various state-of-the-art methods across different datasets detailed in Section~\ref{s:eval}.
We present the results on 1B datasets followed by 100M datasets.

\subsection{Performance on 1B Datasets\label{subsec:1bdataset}}

Figure~\ref{fig:1b_datasets} compares \ours against GGNN and FAISS on billion-size datasets.
\ours outperforms GGNN and FAISS in throughput for compatible 
recalls at 10-recall@10 across all four datasets. 
\revonechanges{
On the SIFT-1B and Deep-1B datasets, \ours{} achieves an average (geomean) throughput improvement of 50.5$\times$ over the competing methods at a high 10-recall@10 of 0.95. 
Specifically, for a 10-recall@10 of 0.9, \ours{} achieves  speedups of 55$\times$, 50$\times$, and 400$\times$ on the SIFT-1B, Deep-1B, and MSSPACEV-1B datasets, respectively. On the more challenging Text-to-Image-1B dataset, \ours and other methods exhibit lower recall values. 
Here, too, \ours{} outperforms the closest baseline by 25$\times$ at a 10-recall@10 of 0.8.
}

To achieve optimal performance in FAISS-PQ, we configure the index values to be large enough to utilise the available GPU memory to the maximum extent possible.
However, FAISS-PQ  fails to achieve higher  recall values (compared to graph-based techniques) because of  losses due to compression. 
The number of distance calculations exponentially increases with recall, as observed in~\cite{ParlayANN}, which negatively impacts its throughput. 
Furthermore, in general, the number of distance computations in FAISS is orders of magnitude higher than in proximity graph-based methods.
\revchanges{
FAISS-SQ achieves higher recall due to milder compression, as each dimension is encoded with a single byte. 
In contrast, FAISS-PQ encodes $M$ dimensions using one byte, resulting in stronger compression but lower recall. However, FAISS-SQ's larger index size requires data swaps between the CPU and GPU, leading to lower throughput compared to FAISS-PQ.
}

For GGNN on a single GPU with 1B dataset, data transfer overhead from CPU to GPU inlined with the search impedes the throughput. 
After utilizing the 80 GB of GPU global memory: the SIFT-1B dataset, for example, requires an additional 123 GB of data to be transferred from the CPU to the GPU during the search. Since the dataset and the graph are each 128 GB and 75 GB (total 203 GB), respectively.
 Transferring this additional 123 GB data over the PCIe bus @ 32 GB/s will take more than 3.8 seconds in practice.
The original GGNN implementation addressed the memory transfer overhead by using eight GPUs to process the shards in parallel.

 
 \ours search runs on GPU but does not require the graph to reside on the GPU. It eliminates the need to transfer the base dataset to GPU by employing compressed vectors (instead of base vectors) present on the GPU.
 Furthermore, the reason for the high throughput of \ours is attributed to the massive parallelism achieved on the GPU (Section~\ref{s:kernels}). 
The re-ranking step, coupled with the integration of Bloom filters to filter processed nodes (prevents duplicate entries in the worklist and allows more deserving nodes to be included in the worklist), significantly contributes to achieving a high recall for \ours.

 
\revchanges{
Figure~\ref{fig:1b_datasets_r100} presents the results of our experiments with 100-recall@100. 
While ANNS evaluations in the literature typically focus on 10-recall@10, real-world applications like recommender systems~\cite{annsinrecommender} often require fetching a larger number of nearest neighbours using ANNS, which are then filtered for final recommendations. 
To reflect this, we extend our evaluation to include 100-recall@100. 
The drop in QPS for \ours is due to the increased worklist length $t$ required to achieve the same recall as 10-recall@10, with the minimum value of $t$ being $k$, where $k$ is 100 for 100-recall@100. 
(The effect of varying worklist length on recall and throughput is detailed in Section~\ref{s:iterationcounts}). 
There is no significant throughput deterioration in FAISS, as the same number of clusters and entries within the posting lists are compared regardless of the recall parameter. 
GGNN, however, exhibits low 100-recall@100 values ($< 0.1$), so we omit its plots in Figure~\ref{fig:1b_datasets_r100}.

}

\begin{table*}[t]
    \centering
    \caption{\revonechanges{ 
    Breakdown of the total ANNS execution time for 10-recall@10 = 0.9 and query batch size = 10,000. 
    }}
 \label{tab:total_execution_time_breakup}
    \begin{tabular}{
    l|rrr|rrr|rrr
    }
         \multicolumn{1}{c|}{\textbf{Activity}} & \multicolumn{3}{c|}{\textbf{BANG}} 
         & \multicolumn{3}{c|}{\textbf{GGNN}}
         & \multicolumn{3}{c}{\textbf{FAISS-SQ}} \\
         \hline
         &\multicolumn{1}{c}{\textbf{SIFT-1B}} &&
         \multicolumn{1}{c|}{\textbf{Deep-1B}}
         &\multicolumn{1}{c}{\textbf{SIFT-1B}} &&
         \multicolumn{1}{c|}{\textbf{Deep-1B}}
         &\multicolumn{1}{c}{\textbf{SIFT-1B}} &&
         \multicolumn{1}{c}{\textbf{Deep-1B}} \\ 
         \cline{2-2} \cline{4-4} \cline{5-5} \cline{7-7} \cline{8-8} \cline{10-10} 
        & Time (ms)  && Time (ms) & Time (ms)  && Time (ms) &Time (ms)  && Time (ms)\\
        \hline
        GPU Processing   & 65.8 && 77.5 & 284.8 && 496.0 & 13.5     && 5.7    \\ 
        CPU Processing   & 15.9 && 20.6 & 0.1   && 0.1 & 14107.0  && 6963.9 \\ 
        Data Transfer  (CPU $\leftrightarrow$ GPU)  & 15.4 && 16.4 &5087.6 && 16303.9& 9.6 && 9.9\\ 
        Overall Processing  & 85.3 && 101.2 & 5372.5 && 16800 & 14130.1 && 6979.5 \\ 
     \end{tabular}
\end{table*}

    
\begin{table}[t]
    \centering
    \caption{\revchanges{
     Percentage breakdown of GPU time (from values in Table~\ref{tab:total_execution_time_breakup}),  and memory usage of \ours kernels.
   }
    }    
    \label{tab:kernel_profiling}
    
 \resizebox{\columnwidth}{!}{
    \begin{tabular}{l@{~}|
    >{\raggedleft\arraybackslash}p{.25cm} 
    >{\raggedleft\arraybackslash}p{0.7cm}| 
    >{\raggedleft\arraybackslash}p{0.25cm} 
    >{\raggedleft\arraybackslash}p{0.7cm}    
    }
         \multicolumn{1}{c|}{\multirow{3}{*}{\textbf{Kernel}}} & \multicolumn{2}{c|}{\textbf{SIFT-1B}}  & \multicolumn{2}{c}{\textbf{Deep-1B}} \\ 
        \cline{2-3}      \cline{4-5} 
        & Time (\%) & Mem (MB)  & Time (\%) & Mem (MB) \\ \hline
        Compute PQ Distance table 
        & 2 & 724 & 2 & 726  \\ 
        Calc neighbours' PQ dist to query point & 46 & 71300  & 49 & 71300  \\ 
        Sort neighbours and merge with Worklist & 18 & 20 & 17 & 21  \\ 
        Filter out visited neighbours & 20 & 3819 & 19 & 3819  \\ 
        Calc exact dist of all candidates (Re-rank) 
        & 4 & 158 & 4 & 448 \\ 
        Pre-fetch next candidate & 10 & 10 & 9 & 18 \\ 
     \end{tabular}
     }
\end{table}

\subsection{Performance on 100M Datasets\label{subsec:100mdataset}}
Figure~\ref{fig:deep100m-sift100m} shows the comparison of throughput and recall on 100M datasets.
As the data structures (i.e., graph and data points) for 100M datasets fit within GPU memory, we report the performance of the two variants of \ours (Section~\ref{s:impl}).
The \banginmemory and  \bangexactdist variants achieve 2$\times$ and 3$\times$ higher throughput compared to the base variant of \ours since there is no CPU-GPU data transfer overhead in the \ours variants. 
Interestingly, the \bangexactdist variant outperforms GGNN. 
We might expect the same performance from the \bangexactdist variant as GGNN, since it uses only GPU-resident data structures as in GGNN and does not use compressed vectors for distance computations to query points. 
However, the main difference between the two is the type of underlying proximity graph.
The k-NN graph structure of GGNN requires more hops or iterations for the search to terminate than \bangexactdist. 
In contrast, in \bangexactdist, the Vamana graph contains long-range edges, reducing the hops, enabling faster convergence with fewer distance computations. 
Note that, compared to \ours{} and FAISS, for GGNN, the performance improvement on 100M datasets over 1B datasets is significant, since the CPU to GPU data transfer overhead, which is present in the 1B datasets, is not present here. 

 \revchanges{We compare with CAGRA on all 100M datasets except Text-to-Image, because the dataset (80 GB) and graph index (12.8 GB), which are required for its execution, exceed the 80 GB global memory capacity of the A100 GPU.  
We observe that CAGRA 
 achieves nearly $1.5\times$--$2\times$ higher throughput than \ours{},
 although it performs $13\times$ more distance computations than \ours{}.  
Furthermore, using the NVIDIA Nsight-Compute profiler (NCU) on the Deep-100M dataset, we measure nearly 7 billion floating-point operations for \ours{} and nearly 12 billion for CAGRA.
However, CAGRA achieves a higher memory throughput of 1270 GB/s compared to 577 GB/s for \ours{}. 
It also has a small memory access overhead due to uncoalesced accesses of 3\% compared to 40\% for 
\ours, 
as measured by NCU using 
the metric
\textit{derived\_\_memory\_l2\_theoretical\_sectors\_global\_excessive}~\cite{nvidia_ncu}.
The higher performance of CAGRA 
is because it uses the highly optimised RAFT~\cite{RAFTLibrary} primitives for bitonic sort/merge and a space-efficient hash table that fits in shared memory, reducing the access latency.
As we target billion-scale datasets, our frequently accessed bloom filter data structure is too large to fit in shared memory; this results in frequent global memory transactions, which impedes performance.
 }

\revonechanges{
\subsection{Search Latency and Memory Usage}
\label{subsec:profiling}
To gain insight into the search latency, we study the time spent by different ANNS methods on (i) CPU execution, (ii) GPU execution, (iii) CPU-GPU data transfer, and (iv) total search time. 
Table~\ref{tab:total_execution_time_breakup} presents the results for \ours and baseline methods on two 1B datasets at a 10-recall@10 value of 0.9. 
For baselines, the overall processing time (the last row) is the sum of the first three components ((i)-(iii)).
However, for \ours, due to its CPU-GPU load balancing with prefetching and asynchronous transfers, the total time is reduced by approx. $12\%$ on both datasets compared to its unoptimised implementation.
Further, the table confirms that GGNN suffers from high transfer overheads due to frequent shard swaps.
FAISS-SQ performs only coarse quantization on the GPU and actual search on the CPU, resulting in low GPU processing time.
FAISS-PQ variant does not reach the target 10-recall@10 of 0.9 and is thus excluded from the table. 
For completeness, FAISS-PQ's execution times for 10-recall@10 of 0.7 on SIFT-1B and DEEP-1B are \{4154, 45.9, 0.06, 4200\} ms and \{4302, 24.9, 0.07, 4327\} ms, respectively, for (i) CPU, (ii) GPU, (iii) transfer, and (iv) total time. 
As these values indicate, unlike FAISS-SQ, the entire search in FAISS-PQ is performed on the GPU.
Furthermore, 
we capture the peak GPU memory usage (in GB) of each method, out of the 80 GB available on the A100 GPU. On SIFT-1B, it is 76.0 (\ours), 78.8 (GGNN), 44.6 (FAISS-PQ), and 2.6 (FAISS-SQ); on Deep-1B, it is 76.3, 68.8, 45.9, and 2.5, respectively. 
The memory usage is dominated by
compressed vectors for \ours, posting lists for GGNN, and graph index shards for FAISS-PQ. 
Thanks to PQ compression, \ours can run on lower-memory GPUs (e.g., 40GB A100 GPU) while maintaining comparable performance (Section~\ref{subsec:chunksize}).
 FAISS-SQ has the least GPU-memory usage, as only centroids reside on the GPU, with search executed on the CPU. 
For \ours, we further measure the distribution of GPU time and memory usage for each kernel. Table~\ref{tab:kernel_profiling} presents the results. 
The majority of the time and memory is utilised by the key kernels involved in distance calculation and neighbour filtering, which is consistent with the analyses in Sections~\ref{ss:dist_calc} and~\ref{ss:bloomfilter}. 

}

\revchanges{
\subsection{Evaluation of Optimisations in {\normalfont \ours}} 
\label{s:ptimization_effects}
Key optimisations in \ours{} include multi-kernel approach, prefetching candidate nodes and asynchronous memory transfers between the host and GPU, discussed in \ref{s:kernels}, \ref{s:prefetching}, and Sections~\ref{ss:data_transfer}, respectively. 
We perform experiments on two popular 1B datasets at a baseline 10-recall@10 value of 0.9 to demonstrate the efficacy of the three important optimisations.
To measure their impact, we disabled each optimisation individually and recorded the throughput. 
The performance improvements for SIFT-1B are 10.1\%, 8\%, and 6.2\%, and for Deep-1B are 17.1\%, 16.5\%, and  13.8\%, respectively. These results demonstrate that the proposed optimisations are significant contributors to \ours{}'s superior performance.

}

\subsection{Effect of Varying Compression Ratio} 
\label{subsec:chunksize}

\begin{figure}[t]
    \centering
    \begin{subfigure}{0.8\columnwidth}
        \centering
    \includegraphics[width=\columnwidth]{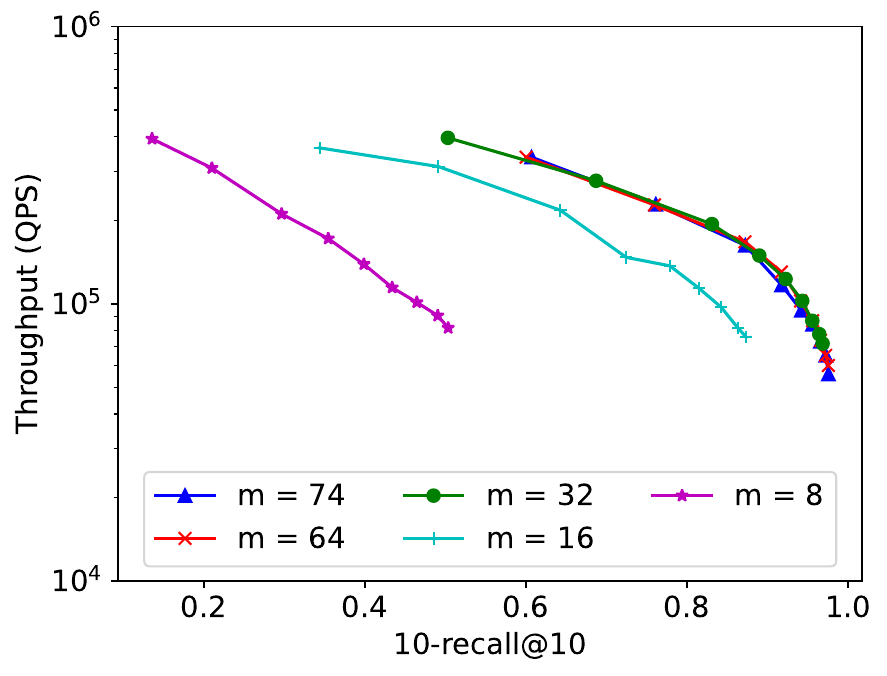} 
    \end{subfigure}
    \hfill
    \caption{
    \revonechanges{
    Throughput (log-scale) vs. 10-recall@10 for \ours on SIFT-1B with varying compressed sizes; $m$ = number of vector subspaces employed in compression (see Section~\ref{s:kernels}).
     }
     } 
    \label{fig:chunk_variation}
\end{figure}

PQ compression techniques invariably introduce a loss in the recall. 
To empirically determine the impact of the compression operation on \ours's recall, we ran \ours search on the SIFT-1B dataset and varied $m$ (i.e. the number of subspaces) during each run while keeping the other configurations the same as in Section~\ref{s:eval}. 
The results are shown in Figure~\ref{fig:chunk_variation}.
 
 While we expect the recall to decrease with high compression (i.e. a lower value of $m$), the result shows no noticeable change in throughput or recall values even when $m$ is decreased from 74 to 32 (representing compression ratios of 0.57 and 0.25, respectively). 
We observed that recall does not drop until a compression ratio of 0.25; below this value, recall begins to decrease. 
Thus, this behaviour allows \ours to operate efficiently on GPUs with limited device memory without compromising recall or throughput. 
For example, we expect \ours search to deliver comparable recall and throughput on an A100 GPU configured with either 
40 GB or 80 GB of device memory. This translates to significant power and cost savings. 

One would expect the throughput to increase with the reduction in compression ratio because it reduces the number of distance computations (requiring fewer PQ distance lookups as the number of subspaces reduces) in the distance calculation kernel. Interestingly, we do not notice this pattern in the results.
The inaccuracies in the distances calculated using compressed vectors increase as $m$ decreases. Hence, \ours's search path needs more hops and detours to converge. 
 The additional time spent in performing more search iterations outweighs 
 the gains from the computations saved on distance calculations, resulting in no significant throughput improvement.

\begin{figure}[t]
    \centering
    \begin{subfigure}{0.8\columnwidth}
        \centering
     \includegraphics[width=\columnwidth]{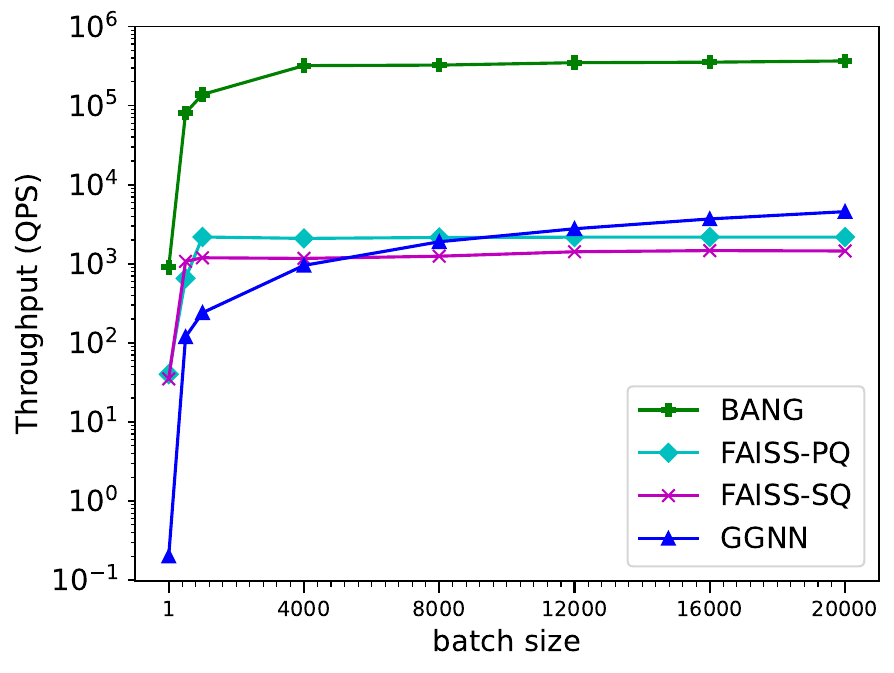}
    \end{subfigure}
       \caption{
    \revonechanges{
     Throughput (log-scale) vs. query batch size on MSSPACEV-1B dataset. Batch size denotes the number of queries processed concurrently.
     }
     } 
    \label{fig:batch_variation}
\end{figure}

\revonechanges{
\subsection{Effect of Varying Query Batch Size} 
\label{subsec:batchsize}
We analyse the impact of input query batch size on throughput by varying it from 1 to 20,000 using the MSSPACEV-1B dataset, which contains more than 10,000 queries.
To include all the baseline algorithms for comparison, we configure the parameters to achieve a 10-recall@10 of $\sim$0.8.
As shown in Figure~\ref{fig:batch_variation}, since \ours and the baselines assign one query per thread block, increasing the batch size allows greater GPU concurrency, improving QPS. 
Throughput increases rapidly for small batch sizes
due to more queries running in parallel on all the available GPU cores but stabilises beyond a saturation point when all GPU cores are fully utilised.  
The saturation points for \ours, FAISS-PQ, and FAISS-SQ occur at batch sizes of approximately 4000, 1000, and 400, respectively. 
\ours’s phased execution strategy enables effective parallelism at higher batch sizes.
}

For GGNN, we observe a relatively higher increase in throughput in relation to the increase in batch size compared to other methods.
As discussed in 
Section~\ref{subsec:profiling}, the data transfer time highly dominates the query processing time. 
Since the data transfer time remains constant and independent of batch size, increasing the batch size leads to a proportionally greater increase in the numerator (number of queries) compared to the denominator (search time), resulting in a relatively higher improvement in throughput for GGNN.



\subsection{\revchanges{Effect of Varying Worklist Size ($t$)}}
\label{s:iterationcounts}
The throughput of \ours{} is inversely proportional to the number of iterations it takes for the search to converge.
We conduct an ablation study to 
investigate
\ours{}'s efficiency in terms of the extra iterations it takes to converge compared to the lower bound on the number of iterations.
The total number of iterations is lower-bounded by 
the size of the worklist (i.e. the search parameter $t$ in Algorithm~\ref{alg:querylookup}) because every entry in the worklist must be visited. 
An increase in $t$ increases the number of iterations required for the search to converge because the number of search paths increases with more entries in the worklist.
We calculate the percentage increase in the number of iterations over the lower bound as $\lambda \coloneq \frac{I-t}{t}\times 100 $, where $I$ is the actual number of iterations taken. 

\begin{table}[t]
    \centering
        \caption{ Variation of \textit{$\lambda$} with respect to search parameter $t$, i.e. the size of the worklist, on SIFT-1B dataset. (smaller $\lambda$ values lead to higher throughputs)}
    \label{tab:iterationcounts}
    \begin{tabular}{r|r|r|r}
         $t$ & $I$& $\lambda$ & 10-recall@10  \\
        \hline
        20 & 62 & 210.0 &0.75 \\
        60 & 104 & 73.3 & 0.91 \\
       100 & 149 & 49.0 &  0.95  \\
        140 & 182 & 30.0 & 0.97 \\
        180 & 222 & 23.3 & 0.98 \\
    \end{tabular}
\end{table}

We vary $t$ from 20 through 180 and measure $\lambda$ and 10-recall@10 for the SIFT-1B dataset with $m=64$ chunks. 
Table~\ref{tab:iterationcounts} presents the results.
Note that recall increases with $t$. So, 
higher values of $t$ increase the recall by enabling the exploration of more search paths.
As we can observe from Table~\ref{tab:iterationcounts},  $\lambda$ decreases with an increase in $t$.
A lower value of $\lambda$ indicates faster convergence with fewer iterations, confirming the efficacy of the recall improvement strategies for the  graph index used. 

Furthermore, in each iteration, multiple queries are processed in parallel, each by a separate thread block. 
The number of queries processed in the first iteration equals the batch size.
In our phased-execution strategy, each query takes one step (a phase) at a time, along with all the other queries in the batch. 
The queries that are converged to termination are excluded from subsequent iterations. 
The total number of iterations ($I$) for the greedy search to complete (i.e., for all queries in a batch to converge) is crucial for throughput; fewer iterations result in higher throughput. 
For a particular $t$ value, we observed that on average, 95\% of queries in a batch are completed in $1.1 \times t$ iterations. 
This demonstrates that most queries take close to the optimal ($t$) number of iterations to complete.
Hence, the parallelisation of \ours does not require sophisticated and heavy-weight thread scheduling mechanisms.

\section{Related Work}
\label{s:related}

 A large body of prior work has tackled the problem of ANNS~\cite{wang2021comprehensive}.
 In this section, we discuss the popular ANNS techniques %
 targeting multicore CPUs, manycore GPUs and custom accelerators.
 We also highlight the key differences between \ours and the existing GPU-based ANNS methods.



\mypara{ANNS on CPU}
Traditionally, ANNS has utilised Tree-based methods (like KD-trees~\cite{kd,kd2} and cover trees) and followed a branch-and-bound approach for search navigation. 
Hashing-based methods are also explored for ANNS. Locality-Sensitive Hash (LSH)~\cite{hash1,hash2} depends on developing hash functions that can generate higher collision probabilities for nearby points than those far apart. 
Both these approaches 
face scalability challenges due to the growing dimensions and size. 

NSW~\cite{nsw} is an early graph-based ANNS method that employs the Delaunay Graph (DG)~\cite{DG} for identifying node neighbours, ensuring near-full connectivity but introducing a high-degree search space with increasing dataset dimensionality.
The graph includes short-range edges for higher accuracy and long-range edges for improved search efficiency, but this results in an imbalance in the graph's degree due to the creation of \textit{traffic hubs}.
HNSW~\cite{hnsw} addresses this by introducing a hierarchical structure to spread neighbours across levels and imposing an upper bound on node degree.
However, scaling HNSW to high dimensions and large datasets continues to be challenging.
NSG \cite{nsg} improves graph-based methods' efficiency and scalability, introducing the Monotonic Relative Neighborhood Graph (MRNG) for reducing index size and search path length while also scaling to billion-scale datasets. 
DiskANN~\cite{diskann} indexes billion-point datasets in hundreds of dimensions through the Vamana graph on commodity hardware (having a 64GB RAM and an inexpensive solid-state drive), leveraging the properties of NSG and NSW. 
It introduces a tunable parameter $\alpha$ for a graph degree-diameter trade-off during construction. It proposes compression schemes to reduce memory consumption and computation costs during the search (see Section~\ref{s:backdiskann}).
SPANN~\cite{SPANN_NEURIPS2021} utilises the inverted index methodology. Unlike other inverted index techniques that use compression techniques (that have some invariable losses) to reduce memory footprint, SPANN uses SSD to store a portion of the index and yet outperforms its predecessors by greatly reducing the number of disk accesses.
HM-ANN~\cite{HMANN_NEURIPS2020} uses memory technologies like Optane PMM to expand the main memory size to fit the graph index and thus avoid compression. 
It extends popular graph-based techniques like HNSW and NSG to run on the heterogeneous memory architecture
with billion-scale datasets.
The survey by Manohar et al.~\cite{dobson2023scaling} extensively studies and compares popular graph-based ANNS algorithms, addressing scalability challenges and proposing parallelization strategies on CPU.

\mypara{ANNS on GPU}
Due to the complexity and throughput requirement of ANNS, offloading these calculations to massively parallel accelerators like GPU has gained considerable attention.  
\revonechanges{
FAISS~\cite{faiss} utilises quantization techniques for dimensionality reduction.  
The underlying search structure of FAISS is the inverted-index (IVF)~\cite{IVF}, where points are typically assigned to buckets using either single-level clustering or two-level clustering.}
FAISS achieves high throughputs on the billion scale but with low recall values. 
PQT~\cite{Wieschollek_2016_CVPR} extends Product Quantization for better GPU performance but has drawbacks, such as longer encoding lengths than PQ-code and high encoding errors.
V-PQT~\cite{CHEN2021107002} addresses this with Vector and Product Quantization Tree, and Parallel-IVFADC~\cite{SOUZA202131} proposes a distributed memory version of FAISS for hybrid CPU-GPU systems scaling up to 256 nodes. 

SONG~\cite{song}, a graph-based ANNS implementation on GPU, offers a notable 
speedup over state-of-the-art CPU counterparts and significant improvement over FAISS 
through 
efficient GPU-centric strategies for ANNS.
GANNS~\cite{GANNS}, built on this foundation, explores enhanced parallelism and occupancy through GPU-friendly data structures and an NSW-based proximity graph construction scheme.
In a related effort, TSDG~\cite{wang2022graphbased} proposes a two-stage graph diversification approach for graph construction and GPU-friendly search procedures catering to various batch query scales.
CAGRA~\cite{cagra}, a recent graph-based ANNS implementation, leverages modern GPU capabilities (e.g., warp splitting)
for substantial performance gains over existing GPU ANNS methods at a million scale.
However, SONG, TSDG, GANNS and CAGRA face limitations in scaling to billion-scale datasets due to the necessity of the entire graph index to reside on the GPU. 

\revonechanges{
Graph-based GPU Nearest Neighbor (GGNN)~\cite{ggnn} search is a recent implementation that outperforms SONG.
It proposes a hierarchical $k$NN graph construction algorithm.
It divides the large input dataset into smaller shards that can fit into GPU memory for performing ANNS on GPU, with the final results merged on the CPU.
GGNN can run on a single GPU at a billion scale but with low throughput. 
To achieve very high throughputs ($> 100,000$) at high recalls ($> 0.95$), it utilises eight GPUs for billion-scale ANNS.
}

We compare our work with algorithms that can handle billion-scale datasets on a single GPU. We find that only FAISS and GGNN fall into this category. 
 Further, unlike these GPU-based approaches,~\ours{} does not require multiple GPUs or data sharding techniques to achieve high recall on billion-scale datasets because it harnesses compressed vectors that can easily fit into a single GPU. 
 Furthermore, instead of running the entire ANNS in one large kernel, \ours explores a \emph{phased-execution} strategy, decomposing the ANNS into distinct phases that can be efficiently executed on CPU/GPU.  Ultimately, \ours delivers significantly higher recall and throughput compared to these approaches.

\mypara{ANNS on other Accelerators} 
Researchers have explored custom hardware, such as FPGAs, for ANNS acceleration, focusing on quantization-based~\cite{FPGA-PQ1, FPGA-PQ2} and graph-based~\cite{FPGA-Graph} approaches. 
However, these methods suffer from frequent data movement between CPU memory and device memory, leading to high energy consumption and lower throughputs.  
To address these limitations, vStore~\cite{FPGA-VSTOR}  proposes an in-storage computing technique for graph-based ANNS within SSDs, avoiding data movement overhead and achieving low search latency with high accuracy.  
On the other hand, TPU-KNN~\cite{chern2022tpuknn} focuses on realizing ANNS on TPUs.
It uses an accurate accelerator performance model considering both memory and instruction bottlenecks. 
\section{Conclusion}
\label{s:concl}


We presented \ours, a novel GPU-based ANNS method that efficiently handles large billion-size datasets exceeding GPU memory capacity, particularly using a single GPU.
It keeps the graph on the CPU and brings together computation on compressed data and optimised GPU parallelisation to achieve high throughput on massive datasets.
\ours enables GPU-CPU computation pipelining and overlapping communication with computation. 
Thus, it is able to optimally utilise 
the 
GPU and CPU resources, and reduce the data transfer over 
the PCIe interconnect between the host and the GPU. 
\revonechanges{
As a result, on billion-scale datasets, it significantly outperforms the state-of-the-art GPU-based methods, achieving \speedupatpointnine 
their throughputs at high recalls.}
\revchanges{Our evaluations on a single GPU show that \ours delivers high throughput with low hardware and operational costs, making it a highly attractive option for practical deployment.}
	\bibliographystyle{IEEEtranS}
	\bibliography{sample}


	\end{document}